\documentclass[sigconf]{acmart}
\usepackage{enumitem}
\usepackage{booktabs} 
\usepackage{graphicx}
\usepackage{subfigure}
\usepackage{bm}
\usepackage{algorithm,algorithmic}
\usepackage{multirow}
\usepackage{mathtools, nccmath}

\DeclarePairedDelimiter{\nint}\lfloor\rceil
\setcopyright{rightsretained}
\allowdisplaybreaks

\acmDOI{10.475/123_4}

\acmISBN{123-4567-24-567/08/06}

\acmConference[SIGIR'19]{SIGIR}{2019}{July 2019, Paris, France}
\acmYear{2019}
\copyrightyear{2019}

\acmArticle{4}
\acmPrice{15.00}

\def \x  {\mathbf{x}}

\def \y  {\mathbf{y}}
\def \u  {\mathbf{u}}
\def \v  {\mathbf{v}}
\def \w  {\mathbf{w}}
\def \R  {\mathbf{R}}
\def \U  {\mathbf{U}}
\def \V  {\mathbf{V}}

\def \b  {\mathbf{b}}
\def \d  {\mathbf{d}}
\def \D  {\mathbf{D}}
\def \B  {\mathbf{B}}
\def \I  {\mathbf{I}}
\def \X  {\mathbf{X}}
\def \Y  {\mathbf{Y}}
\def \P  {\mathbf{P}}
\def \Q  {\mathbf{Q}}
\def \R  {\mathbf{R}}
\def \bSigma {\bm{\Sigma}}

\def \bbeta {\bm{\eta}}
\def \bxi {\bm{\xi}}

\begin{document}
\title{Compositional Coding for Collaborative Filtering}
\titlenote{Jianling Sun is the corresponding author.}
\author{Chenghao Liu$^{1,2}$, Tao Lu$^{2,4}$, Xin Wang$^3$, Zhiyong Cheng$^{5,6}$, Jianling Sun$^{2,4}$, Steven C.H. Hoi$^1$}
\affiliation{\institution{$^1$Singapore Management University, $^2$Zhejiang University, $^3$Tsinghua University}\institution{$^4$Alibaba-Zhejiang University Joint Institute of Frontier Technologies}\institution{$^5$Shandong Computer Science Center (National Supercomputer Center in Jinan)}\institution{$^6$Qilu University of Technology (Shandong Academy of Sciences)}}
\email{{twinsken, 3140102441, sunjl}@zju.edu.cn, xin_wang@tsinghua.edu.cn, jason.zy.cheng@gmail.com, chhoi@smu.edu.sg}



\renewcommand{\shortauthors}{Chenghao and Tao, et al.}
\begin{abstract}
Efficiency is crucial to the online recommender systems, especially for the ones which needs to deal with tens of millions of users and items. Because representing users and items as binary vectors for Collaborative Filtering (CF) can achieve fast user-item affinity computation in the Hamming space, in recent years, we have witnessed an emerging research effort in exploiting binary hashing techniques for CF methods. However, CF with binary codes naturally suffers from low accuracy due to limited representation capability in each bit, which impedes it from modeling complex structure of the data. 

In this work, we attempt to improve the efficiency without hurting the model performance by utilizing both the accuracy of real-valued vectors and the efficiency of binary codes to represent users/items. In particular, we propose the Compositional Coding for Collaborative Filtering (CCCF) framework, which not only gains better recommendation efficiency than the state-of-the-art binarized CF approaches but also achieves even higher accuracy than the real-valued CF method. Specifically, CCCF innovatively represents each user/item with a set of binary vectors, which are associated with a sparse real-value weight vector. Each value of the weight vector encodes the importance of the corresponding binary vector to the user/item. The continuous weight vectors greatly enhances the representation capability of binary codes, and its sparsity guarantees the processing speed. Furthermore, an integer weight approximation scheme is proposed to further accelerate the speed. Based on the CCCF framework, we design an efficient discrete optimization algorithm to learn its parameters. Extensive experiments on three real-world datasets show that our method outperforms the state-of-the-art binarized CF methods (even achieves better performance than the real-valued CF method) by a large margin in terms of both recommendation accuracy and efficiency. We publish our project at https://github.com/3140102441/CCCF.

\end{abstract}

\ccsdesc[500]{Information systems~Recommender System}
\ccsdesc[500]{Humancentered computing~Collaborative filtering}

\keywords{Recommendation, Collaborative Filtering, Discrete Hashing}
\maketitle

\section{Introduction}
Real-world recommender systems often have to deal with large numbers of users and items especially for online applications, such as e-commerce or music streaming services \cite{liu2016online,liu2017collaborative,wang2017interactive,cheng2018aspect,cheng2017exploiting}. For many modern recommender systems, a de facto solution is often based on Collaborative Filtering (CF) techniques, as exemplified by Matrix Factorization (MF) algorithms \cite{koren2009matrix}. The principle of MF is to represent users' preferences and items' characteristics into $r$ low-dimensional vectors,  based on the $m\times n$ user-item interaction matrix of $m$ users and $n$ items. With the obtained user and item vectors (in the offline training stage), during the online recommendation stage, the preference of a user towards an item  is computed by the dot product of their represented vectors. However, when dealing with large numbers of users and items (e.g., millions or even billions of users and items), a naive implementation of typical collaborative filtering techniques (e.g., based on MF)  will lead to very high computation cost for generating preferred item ranking list for a target user \cite{li2017neural}. Specifically, recommending the top-$k$ preferred items for a user from those $n$ items costs $O(nr+n\log k)$ with real-valued vectors. As a result, this process will become a critical efficiency bottleneck in practice where the recommender systems typically require a real-time response for large-scale users simultaneously. Therefore, a fast and scalable yet accurate CF solution is crucial towards building real-time recommender systems.

Recent years have witnessed extensive research efforts for improving the efficiency of CF methods for scalable recommender systems. One promising paradigm is to explore the hashing techniques \cite{zhou2012learning,liu2014collaborative,zhang2014preference} to represent users/items with binary codes instead of the real-value latent factors in traditional MF methods. In this way, the dot-products of user vector and item vector in MF can be completed by fast bit-operations in the Hamming space \cite{zhou2012learning}. Furthermore, by exploiting special data structures for indexing all items, the computational complexity of generating top-K preferred items is sub-linear or even constant \cite{wang2012semi,zhang2014preference}, which significantly accelerates the recommendation process. 

However, learning the binary codes is generally NP-hard \cite{haastad2001some} due to its discrete constraints. Given this NP-hardness, a two-stage optimization procedure \cite{zhou2012learning,liu2014collaborative,zhang2014preference}, which first solves a relaxed optimization problem through ignoring the discrete constraints and then binarizes the results by thresholding, becomes a compromising solution. Nevertheless, this solution suffers from a large quantization loss \cite{zhang2016discrete} and thus fails to preserve the original data geometry (user-item relevance and user-user relationship) in the continuous real-valued vector space. As accuracy is arguably the most important evaluation metric for recommender systems, researchers put lots of efforts to reduce the quantization loss by direct discrete optimization \cite{zhang2016discrete,lian2017discrete,liu2018discrete}. In spite of the advantages of this improved optimization method, compared to real-valued vectors, CF with binary codes naturally suffers from low accuracy due to limited representation capability in each bit, which impedes it from modeling complex relationship between users and items.
\begin{figure}[!htb]
  \centering
    \includegraphics[width=0.45\textwidth]{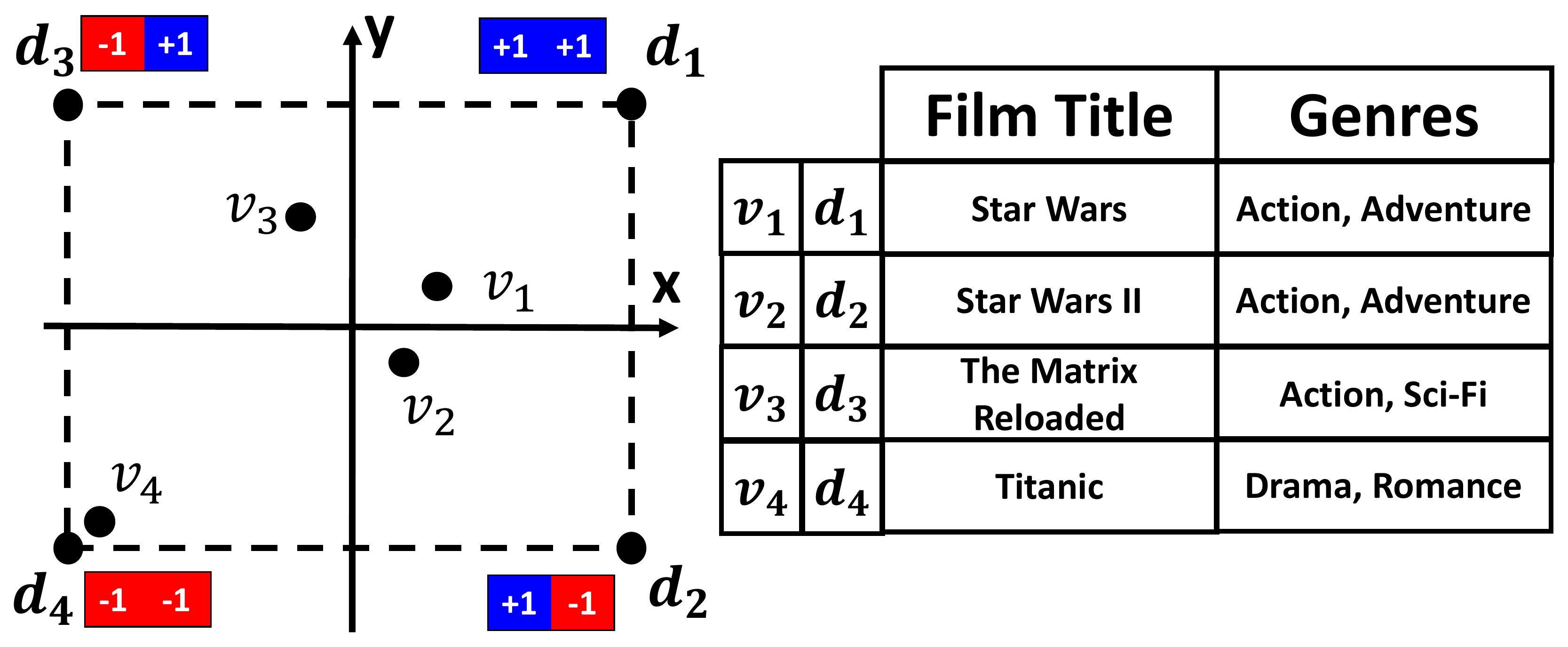}
    \vspace{-0.1cm}
  \caption{A toy example to illustrate the limitation of binary codes of Discrete Collaborative Filtering (DCF) \cite{zhang2016discrete}. $\v_1,\v_2,\v_3$, and $\v_4$ denote the real-valued vectors for item embeddings, and $\d_1,\d_2,\d_3$, and $\d_4$ denote the binary codes for item embeddings. According to the film title and genres, $v_1$ is the most similar to $v_2$, followed by $v_3$, while they are all dissimilar to $v_4$. However, the binary codes learned by DCF cannot preserve the intrinsic similarity due to the limited representation capability of binary codes.}
\label{showcase}
\vspace{-0.1cm}
\end{figure}

Figure \ref{showcase} gives an example to illustrate the limit of binary codes. From the ``film title" and ``genres", we can see that \textit{Star Wars} is the most similar with \textit{Star Wars II}, followed by \textit{The Matrix Reloaded}, and all of them are action movies while \textit{Titanic} is remarkably dissimilar to them which is categorized as a {\it Drama} movie. The real-valued vectors could easily preserve the original data geometry in the continuous vector space (e.g., intrinsic movie relationships), like $\v_1,\v_2,\v_3$, and $\v_4$. However, if we preserve the geometric relations of \textit{Star Wars}, \textit{Star Wars II} and \textit{The Matrix Reloaded} by representing them using the binary codes $\d_1$, $\d_2$, and $\d_3$ in the Hamming space, the binary code of movie \textit{Titanic} $\d_4$ will become close to $\d_2$ and $\d_3$, which unfortunately leads to a large error.

In this work, we attempt to improve the efficiency without hurting the model performance. We propose a new user/item representation named ``Compositional Coding", which utilizes both the accuracy of real-valued vectors and the efficiency of binary codes to represent users and items. To improve the representation capability of the binary codes, each user/item is represented by $G$ components of $r$-dimensional binary codes ($r$ is relatively small) together with a $G$-dimensional sparse weight vector.  The weight vector is real-valued and indicates the importance of the corresponding component of the binary codes. Compared to the binary codes with same length (equal to $Gr$), the real-valued weight vector significantly enriches the representation capability. Meanwhile, the sparsity of the weight vector could preserve the high efficiency.  To demonstrate how it works, we derive the Compositional Coding for Collaborative Filtering (CCCF) framework. To tackle the intractable discrete optimization of CCCF, we develop an efficient alternating optimization method which iteratively solves mixed-integer programming subproblems. Besides, we develop an integer approximation strategy for the weight vectors. This strategy can further accelerate the recommendation speed. We conduct extensive experiments in which our promising results show that the proposed CCCF method not only improves the accuracy but also boosts retrieval efficiency over state-of-the-art binary coding methods.

\section{Preliminaries}
In this section, we first review the two-stage hashing method for collaborative filtering.  Then, we introduce the direct discrete optimization method, which has been used in Discrete Collaborative Filtering (DCF) \cite{zhang2016discrete}. Finally, we discuss the limitation of binary codes in representation capability.

\subsection{Two-stage Hashing Method}
Matrix Factorization (MF) is the most successful and widely used CF based recommendation method. It represents users and items with real-valued vectors. Then an interaction of the corresponding user and item can be efficiently estimated by the inner product. Formally, given a user-item interaction matrix $\R \in \mathbb{R}^{m\times n}$ with $m$ users and $n$ items.  Let $\u_i \in \mathbb{R}^r$  and  $\v_j\in \mathbb{R}^r$ denote the latent vector for user $i$ and item $j$ respectively. Then, the predicted preference of user $i$ towards item $j$ is formulated as
$\hat{r}_{ij}=\u_i^\top\v_j.$
To learn all user latent vectors  $\U=[\u_1,\dots,\u_m]^\top\in\mathbb{R}^{m\times r}$ and item latent vectors $\V=[\v_1,\dots,\u_n]^\top\in\mathbb{R}^{n\times r}$, MF minimizes the following regularized squared loss on the observed ratings:
\begin{align}
\arg\min\limits_{\U,\V}\sum_{(i,j)\in\mathcal{V}}(R_{ij}-\u_i^\top\v_j)^2+\lambda R(\U,\V),
\end{align}
where $\mathcal{V}$ denotes the all observed use-item pairs and $R(\U,\V)$ is the regularization term with respect to $\U$ and $\V$ controlled by $\lambda >0$.  To improve recommendation efficiency, after we obtain the optimized user/item real-valued latent vectors, the two-stage hashing method use binary quantization (rounding off \cite{zhang2014preference} or rotate  \cite{zhou2012learning,liu2014collaborative}) to convert the continuous latent representations into binary codes. Let us denote  $\B=[\b_1,\dots,\b_m]^\top\in\{\pm 1\}^{m\times r}$ and $\D=[\d_1,\dots,\d_n]^\top\in\{\pm 1\}^{n\times r}$ respectively as $r$-length user/item binary codes, then the inner product between the binary codes of user and item can be formulated as $\b_i^\top\d_j=2H(\b_i,\d_j)-r,$where $H(\cdot)$ denotes the Hamming similarity. Based on fast bit operations, Hamming distance computation is extremely efficient. However, this method usually incurs a large quantization error since the binary bits are obtained by thresholding real values to integers, and thus it cannot preserve the original data geometry in the continuous vector space \cite{zhang2016discrete}.

\subsection{Direct Discrete Optimization Method}
To circumvent the above issues, direct discrete optimization method has been proposed in Discrete Collaborative Filtering (DCF) \cite{zhang2016discrete} and its extension \cite{lian2017discrete,zhang2017discrete,liu2018discrete}. More formally, it learns the binary codes by optimizing the following objective function:
\begin{align}
&\arg\min\limits_{\B,\D}\sum_{(i,j)\in\mathcal{V}}(R_{ij}-\b_i^\top\d_j)^2+\lambda R(\B,\D)\nonumber\\
&s.t. \quad \B\in\{\pm 1\}^{m\times r},\D\in\{\pm 1\}^{n\times r}.
\end{align}
This objective function is similar to that of a conventional MF task, except the discrete constraint on the user/item embeddings. By additionally imposing balanced and de-correlated constraints, it could derive compact yet informative binary codes for both users and items.

However, compared to real-valued vectors, CF with binary codes naturally suffers from low accuracy due to limited representation capability in each bit. Specifically, the $r$-dimensional real-valued vector space have infinite possibility to model users and items, while the number of unique binary codes in Hamming space is $2^r$. An alternative way to resolve this issue is to use a longer code. Unfortunately, it will adversely hurt the generalization of the model especially in the sparse scenarios.

\begin{figure}[h]
  \centering
    \includegraphics[width=0.48\textwidth]{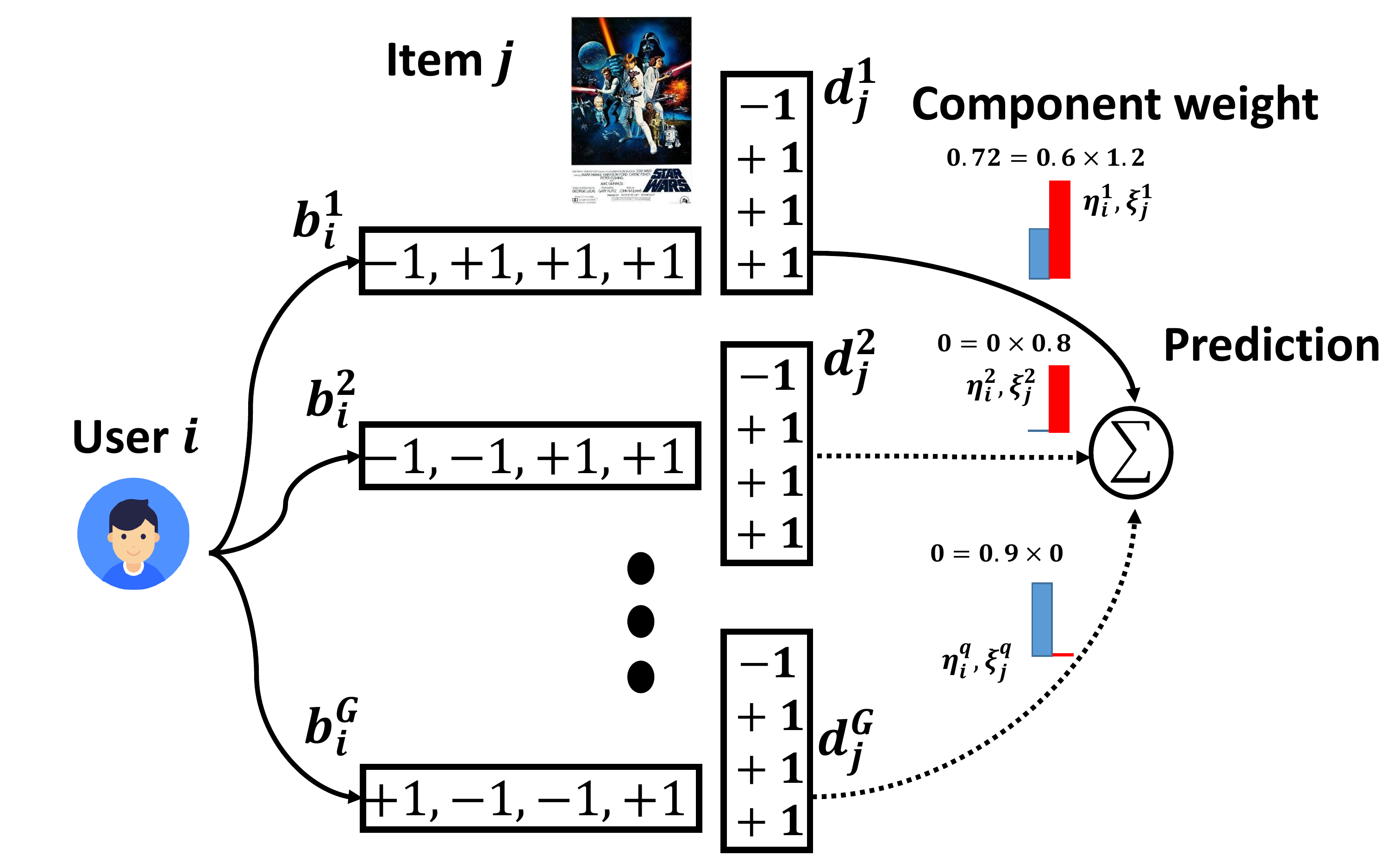}
  \caption{Illustration of how the compositional coding framework makes a prediction for user $i$ given item $j$.}
\label{showcase2}
\end{figure}

\section{Compositional Coding for Collaborative Filtering}
\label{cccf_sec}
\subsection{Intuition}
In this work, we attempt to improve the efficiency of CF without hurting the prediction performance. We propose a new user/item representation named ``Compositional Coding", which utilizes both the accuracy of real-valued vectors and the efficiency of binary codes to represent users and items. Since the fixed distance between each pair of binary codes impedes them from modeling different magnitudes of relationships between users and items, we assign real-valued weight to each bit to remarkably enrich their representation capability. Besides, the importance parameters could implicitly prune unimportant bits by setting their importance parameters close to $0$, which naturally reduces the computation cost.

\vspace{-0.1cm}
\subsection{Overview}
In general, the proposed compositional coding framework assumes each user or item is represented by $G$ components of $r$-dimensional binary codes ($r$ is relatively small) and one $G$-dimensional sparse weight vector.
Formally, denote by $$\b^{(1)}_i,\dots,\b^{(G)}_i \in \{\pm 1\}^{r}, \bbeta_i=\Big(\eta_i^{(1)},\cdots,\eta_i^{(G)}\Big) \in \mathbb{R}^G$$ the compositional codes for the $i$-th user, and  $$\d^{(1)}_j,\dots,\d^{(G)}_j\in \{\pm 1\}^{r}, \bxi_j=\Big(\xi_j^{(1)},\cdots,\xi_j^{(G)}\Big) \in \mathbb{R}^G$$ the compositional codes for the $j$-th item,  respectively. Then the predicted preference of user $i$ for item $j$ is computed by taking the weighted sum of the inner product with respect to each of the $G$ components:
\begin{align}
\hat{r}_{ij} = \sum^G_{k=1}w^{(k)}_{ij}(\b_i^{(k)})^\top\d_j^{(k)},
\end{align}
where $w_{ij}^{(k)}=\eta_i^{(k)}\cdot\xi_j^{(k)}$ is the importance weight of the $k$-th component of binary codes with respect to user $i$ and item $j$.

Figure \ref{showcase2} illustrates how to estimate a user-item interaction using compositional codes. The inner product of each component of binary codes $(\b_i^{(k)})^\top\d_j^{(k)}$ can be efficiently computed in Hamming space using fast bit operations. The importance weight of the $k$-th component of binary codes  $w_{ij}^{(k)}$ is obtained through a multiplication operation over $\eta_i^{(k)}$ and $\xi_j^{(k)}$, which ensures that the importance weight will be assigned a high value if and only if both user and item weights are large and will become zero if either of them is zero. It is worth noting that we use the multiplication instead of addition operation over the user and item weight to achieve the high sparsity of the sparse weight vector, which can lead to a significant reduction of computation cost during online recommendation.

Compared to representing users/items with binary codes, the key advantage of the proposed framework is that the sparse real-valued weight vector substantially increases the representation capacity of user/item embeddings by the compositional coding scheme. Recalling the example shown in Figure \ref{showcase}, we can preserve the movie relations by assigning them with different weight vectors, so as to predict user-item interactions with a lower error. Another benefit is mainly from the idea of compositional matrix approximation \cite{lee2013local,zhang2013improve}, which is more suitable for real-world recommender systems, since the interaction matrix is very large and composed of diverse interaction behaviours. Under this view, the proposed compositional coding framework is characterized by multiple components of binary codes. Each component of binary codes can be employed to discover the localized relationships among certain types of similar users and items and thus can be more accurate in this particular region (e.g., young people viewing action movies while old people viewing romance movies). Therefore, the proposed compositional coding framework can encode users and items in a more informative way.

In addition to the improved representation and better accuracy, the compositional coding framework does not lose the high efficiency advantage of binary codes, and can even gain more efficiency when imposing sufficient sparsity. In order to show this, we analyze the time cost of the proposed framework and compare it against the conventional binary coding framework. In particular, assume the time cost of calculating the Hamming distance of $r$-bit codes is $T_{h}$ and the time cost of weighted summation of all of inner product results is $T_{s}$, then the total cost for a user-item similarity search is
\begin{align}
nnz(\w)\times T_{h}+T_{s}, \nonumber
\end{align}
where $nnz(\w)$ denotes the average number of non-zero values in component weight vector $\w\in\mathbb{R}^G$, which is much smaller than $G$ in our approach due to the sparsity of the user and item weights. Similarly, the cost by conventional hashing based CF models using $(G\times r)$-bit binary codes is $G\times T_{h}$ which can be higher than ours.

\textbf{Remark.} The proposed compositional coding framework is rather general. When identical weight vectors are used, it is degraded to the binary codes of conventional hashing based CF methods. Compositional codes could also be regarded as real-valued latent vectors of CF model, if we adopt the identical binary codes for each component. In summary, compositional code is a flexible representation which could benefit from both the strong representation capability of real-valued vectors and the fast similarity search of binary codes.

\vspace{-0.2cm}
\subsection{Formulation}
In this section, we instantiate the compositional coding framework on Matrix Factorization (MF) models and term the proposed method as the Compositional Coding for Collaborative Filtering (CCCF). Note that the proposed compositional coding framework can be potentially applicable to other types of CF models beyond MF.

Follow the intuition of compositional coding, we assume that there exists a distance function $d$ that measures distances in the space of users ($i=1,\dots,m$) or items ($j=1,\dots,n$). The distance function leads to the notion of neighborhoods of user-user and item-item pairs. Its assumption states that the rating of user-item pair $(i,j)$ could be approximated particularly well in its neighbourhood. Thus, we identify $G$ components surrounding $K$ anchor points $(i^\prime_1,j^\prime_1),\dots, (i^\prime_G,j^\prime_G)$. Follow the setting of compositional matrix approximation \cite{lee2013local,lee2014local},  user weight $\eta^{(k)}_i$ can be instantiated with an Epanechnikov kernel\footnote{Similar to \cite{lee2014local}, we also tried uniform
and triangular kernel, but the performance was worse than Epanechnikov kernel, in agreement with the theory of kernel smoothing \cite{wand1994kernel}.} \cite{wand1994kernel}, which is formulated as:
\begin{small}
\begin{align}
\label{weight}
\eta^{(k)}_i=\frac{3}{4}\Big(1-d(i,i^\prime_t)^2\Big)\textbf{1}\big[d(i,i^\prime_t)<h\big],
\end{align}
\end{small}
where $h>0$ is a bandwidth parameter, $\bf{1}[\cdot]$ is the indicator function and $d(\cdot)$ is a distance function to measure the similarity between $i$ and $i^\prime_t$ (we will discuss it in Section \ref{distance}).  A large value of $h$ implies that $\bbeta_i$ has a wide spread, which means most of the user component weights are non-zero. In contrast, a small $h$ corresponds to narrow spread of $\bbeta_i$ and most of the user components will be zero. Item weight $\xi^{(k)}_j$ follows the analogous formulation. In this way, we could endow the weight vector $w^{(k)}_{ij}$ for user-item pair, which is defined as multiplication over user and item weight, with the sparsity. The method for selecting anchor points for each component is important as it may affect the values of component weights and further affect the generalization performance. A natural way is to uniformly sample them from the training set. In this work, we run $k$-means clustering with the user/item latent vectors and use the $G$ centroids as anchor points.

To learn the compositional codes, we adopt the squared loss to measure the reconstruction error as the standard MF method. For each set of binary codes, in order to maximize the entropy of each bit and make it as independent as possible, we impose balanced partition and decorrelated constraints. In summary, learning the compositional codes for users and items can be formulated into the following optimization task:
\begin{small}
\begin{align}
\label{problem}
\min &\sum_{(i,j)\in\mathcal{V}}\Big(R_{ij}-\sum^G_{k=1}w_{ij}^{(k)}(\b_i^{(k)})^\top\d_j^{(k)}\Big)^2\nonumber\\
\text{s.t.}  & \underbrace{\textbf{1}_m\B^{(k)}=0,\textbf{1}_n\D^{(k)}=0}_{\text{balanced partition}},\underbrace{(\B^{(k)})^\top\B^{(k)}=m\I_r,(\D^{(k)})^\top\D^{(k)}=n\I_r}_{\text{decorrelation}} \nonumber\\
&\B^{(k)} \in \{\pm 1\}^{m\times r},\; \D^{(k)}\in \{\pm 1\}^{n\times r}, \quad k=1,\dots,G.
\end{align}
\end{small}
where we denote $\B^{(k)}=[\b^{(k)}_1,\dots,\b^{(k)}_m]^\top\in\{\pm 1\}^{m\times r}$ and $\D^{(k)}=[\d^{(k)}_1,\dots,\d^{(k)}_n]^\top\in\{\pm 1\}^{n\times r}$ respectively as user and item binary codes in the $k$-th set of binary codes. The problem formulated in (\ref{problem}) is a mixed-binary-integer program which is a challenging task since it is generally NP-hard and involves a combinatorial search over $O(2^{Gr(m+n)})$. An alternative way is to impose auxiliary continuous variables $\X^{(k)} \in \mathcal{B}$ and $\Y^{(k)} \in \mathcal{D}$ for each set of binary codes, where
$\mathcal{B}^{(k)}=\{\X^{(k)}\in\mathcal{R}^{m\times r}|\textbf{1}_m\X^{(k)}=0,(\X^{(k)})^\top\X^{(k)}=m\I_r\}$ and $\mathcal{D}^{(k)}=\{\Y^{(k)}\in\mathcal{R}^{n\times r}|\textbf{1}_n\Y^{(k)}=0,(\Y^{(k)})^\top\Y^{(k)}=n\I_r\}$. Then the balanced and de-correlated constraints can be softened by $\min_{\X^{(k)}\in\mathcal{B}^{(k)}}\|\B^{(k)}-\X^{(k)}\|_F$ and $\min_{\Y^{(k)}\in\mathcal{D}^{(k)}}\|\D^{(k)}-\Y^{(k)}\|_F$, respectively. Finally, we can solve problem (\ref{problem}) with respect to the $k$-the set of codes in a computationally tractable manner:
\begin{small}
\begin{align}
\label{problem2}
\min\limits_{\B^{(k)},\D^{(k)},\X^{(k)},\Y^{(k)}}&\sum_{(i,j)\in\mathcal{V}}\Big(R_{ij}-\sum^G_{k=1}w_{ij}^{(k)}(\b_i^{(k)})^\top\d_j^{(k)}\Big)^2\nonumber\\
&\quad+\alpha_1\|\B^{(k)}-\X^{(k)}\|_F+\alpha_2\|\D^{(k)}-\Y^{(k)}\|_F\nonumber\\
\text{s.t.}\quad\quad\quad&\quad\B^{(k)} \in \{\pm 1\}^{m\times r},\D^{(k)}\in \{\pm 1\}^{n\times r},
\end{align}
\end{small}
where $\alpha_1$ and $\alpha_2$ are tuning parameters. Since
$\text{tr}((\B^{(k)})^\top\B^{(k)})=\text{tr}((\X^{(k)})^\top\X^{(k)})=mr,$ and $\text{tr}((\D^{(k)})^\top\D^{(k)})=\text{tr}((\Y^{(k)})^\top\Y^{(k)})=nr,$ the above optimization task can be turned into the following
\begin{small}
\begin{align}
\label{problem3}
&\min\limits_{\B^{(k)},\D^{(k)},\X^{(k)},\Y^{(k)}}\sum_{(i,j)\in\mathcal{V}}(R_{ij}-\sum^G_{k=1}w_{ij}^{(k)}(\b_i^{(k)})^\top\d_j^{(k)})^2\nonumber\\
&\quad\quad\quad-2\alpha_1\text{tr}((\B^{(k)})^\top\X^{(k)})-2\alpha_2\text{tr}((\D^{(k)})^\top\Y^{(k)})\nonumber\\
\text{s.t.}  \quad & \textbf{1}_m\X^{(k)}=0,\textbf{1}_n\X^{(k)}=0,(\Y^{(k)})^\top\Y^{(k)}=m\I_r,(\Y^{(k)})^\top\Y^{(k)}=n\I_r \nonumber\\
&\B^{(k)} \in \{\pm 1\}^{m\times r},\D^{(k)}\in \{\pm 1\}^{n\times r},\quad\quad
\end{align}
\end{small}
which is the proposed learning model for CCCF. Note that we do not discard the binary constraints but directly optimize the binary codes of each component. Through joint optimization
for the binary codes and the auxiliary real-valued variables, we can achieve nearly balanced and un-correlated binary codes. Next, we will introduce an efficient solution for the mixed-integer
optimization problem in Eq. (\ref{problem3}).

\subsection{Optimization}
We employ alternative optimization strategy to solve the above problem. Each iteration alternatively updates
$\B^{(k)}$, $\D^{(k)}$, $\X^{(k)}$ and $\Y^{(k)}$. The details are given below.

\textbf{Learning $\B^{(k)}$ and $\D^{(k)}$:}  In this subproblem, we update $\B^{(k)}$ with fixed $\D^{(k)}$, $\X^{(k)}$ and $\Y^{(k)}$. Since the objective function in Eq. (\ref{problem3}) is based on summing over users of each component independently, so we can update binary codes for each user and item in parallel. Specifically, learning binary codes for user $i$ with respect to component $k$ is to solve the following optimization problem:
\begin{small}
\begin{align}
\min_{\b_i^{(k)}\in\{\pm 1\}^r}&(\b^{(k)}_i)^\top\Bigg(\sum_{j\in\mathcal{V}_i}(w_{ij}^{(k)})^2\d_j^{(k)}(\d_j^{(k)})^\top\Bigg)\b^{(k)}_i
\nonumber\\&-2\Big(\sum_{j\in\mathcal{V}_i}\tilde{r}_{ij}(\d_j^{(k)})^\top\Big)\b_i^{(k)} -2\alpha_1(\x^{(k)}_i)^\top\b^{(k)}_i,
\end{align}
\end{small}
where $\tilde{r}_{ij}= r_{ij}-\sum_{\tilde{k}\neq k }w_{ij}^{\tilde{k}}(\b_i^{\tilde{k}})^\top\d_j^{\tilde{k}}$ is the residual of observed rating excluding the inner product of component $k$.

Due to the discrete constraints, the optimization is generally NP-hard, we adopt the bitwise learning method called Discrete Coordinate Descent \cite{shen2015supervised,shen2017classification} to update $\b^{(k)}_i$. In particular, denoting $b^{(k)}_{iq}$  as the $q$th bit of $\b^{(k)}_i$ and $\b^{(k)}_{i\bar{q}}$ as the rest codes excluding $b^{(k)}_{iq}$, DCD update $b^{(k)}_{iq}$ while fixing $\b^{(k)}_{i\bar{q}}$. Thus, the updating rule for user binary code $b^{(k)}_{iq}$ can be formulated as
\begin{align}
\label{b_square}
b^{(k)}_{iq} \leftarrow sgn(O(-\hat{b}^{(k)}_{ik},b^{(k)}_{iq}))
\end{align}
where $\hat{b}^{(k)}_{iq}=\sum_{j\in\mathcal{V}_i}(\tilde{r}_{ij}-(w_{ij}^{(k)})^2(\d^{(k)}_{j\bar{q}})^\top\b^{(k)}_{i\bar{q}})d^{(k)}_{jq}+\alpha_1 x^{(k)}_{iq}$, $\d^{(k)}_{j\bar{q}}$ is the rest set of item codes excluding $d_{jk}$
and $O(x,y)$ is a function that $O(x,y)=x$ if $x\neq0$ and $O(x,y) =y$ otherwise. We iteratively update each bit until the procedure convergence. Note that the computational complexity of updating $\B^{(k)}$ is $O\big(\#iter (mnr^2)\big)$ which is a critical efficiency bottleneck when $m$ or $n$ is large. To efficiently compute $\hat{b}^{(k)}_{iq}$, we rewrite $\hat{b}^{(k)}_{iq}$ as
\begin{small}
\begin{align}
\hat{b}^{(k)}_{iq}=\sum_{j\in\mathcal{V}_i}\tilde{r}_{ij}d^{(k)}_{jq}-\sum_{j\in\mathcal{V}_i}(w_{ij}^{(k)})^2(\d^{(k)}_j)^\top\b_i^{(k)} d^{(k)}_{jq} + \sum_{j\in\mathcal{V}_i}b^{(k)}_{iq} + \alpha_1 x^{(k)}_{iq}, \nonumber
\end{align}
\end{small}
which reduces the computational cost to $O\big(\#iter (m+n)r^2\big)$.

Similarly, we could learn binary code for item $j$ in component $k$ by solving
\begin{small}
\begin{align}
\min_{\d_j^{(k)}\in\{\pm 1\}^r}&(\d^{(k)}_j)^\top(\sum_{i\in\mathcal{V}_j}(w_{ij}^{(k)})^2\b_i^{(k)}(\b_i^{(k)})^\top)\d^{(k)}_j\nonumber\\
&-2(\sum_{i\in\mathcal{V}_j}\tilde{r}_{ij}(\b_i^{(k)})^\top)\d_j^{(k)}-2\alpha_2(\y^{(k)}_j)^\top\d^{(k)}_j.\nonumber
\end{align}
\end{small}
Denote $d^{(k)}_{jq}$ as the $q$-th bit of $\d^{(k)}_j$ and $\d^{(k)}_{j\bar{q}}$ as the rest codes excluding $d^{(k)}_{jq}$, we update each bit of $\d^{(k)}_j$ according to
\begin{align}
\label{d_square}
d^{(k)}_{jq} \leftarrow sgn(O(-\hat{d}^{(k)}_{jq},d^{(k)}_{jq}))
\end{align}
where $\hat{d}^{(k)}_{jq}=\sum_{i\in\mathcal{V}_j}(\tilde{r}_{ij}-(w_{ij}^{(k)})^2(\b^{(k)}_{i\bar{q}})^\top\d^{(k)}_{j\bar{q}})b^{(k)}_{iq}+\alpha_2 y^{(k)}_{jq}$.

\textbf{Learning $\X^{(k)}$ and $\Y^{(k)}$:} When fixing $\B^{(k)}$, learning $\X^{(k)}$ could be solved via optimizing the following objective function:
\begin{align}
\max_{\X^{(k)}} tr(\B^{(k)}(\X^{(k)})^\top), \quad \textbf{1}_m^\top\X^{(k)}=0, (\X^{(k)})^\top\X^{(k)}=m\I_r. \nonumber
\end{align}
It can be solved by the aid of SVD according to \cite{liu2014discrete}. Let $\bar{\B}^{(k)}$ be a column-wise zero-mean matrix, where $\bar{B}^{(k)}_{ij}=B^{(k)}_{ij}-\frac{1}{m}\sum_iB^{(k)}_{ij}$. Assuming $\bar{\B}^{(k)}=\P^{(k)}_b\bSigma^{(k)}_b(\Q^{(k)}_b)^\top$ as its SVD, where each column of $\P^{(k)}_b \in \mathbb{R}^{m\times r^\prime}$ and $\Q^{(k)}_b \in \mathbb{R}^{r\times r^\prime}$ represents the left and right singular vectors corresponding to $r^\prime$ non-zero singular values in the diagonal matrix $\bSigma^{(k)}_b$. Since $\bar{\B}^{(k)}$ and $\Q^{(k)}_b$ have the same row, we have $\textbf{1}^\top{\P^{(k)}}_b=0$ due to $\textbf{1}^\top\bar{\B}^{(k)}=0$. Then we construct matrices $\hat{\P}^{(k)}_b$ of size $m\times (r-r^\prime)$ and $\hat{\Q}^{(k)}_b$ of size $r\times (r-r^\prime)$ by employing a Gram-Schmidt process such that $(\ \hat{\P}^{(k)})_b^\top\hat{\P}^{(k)}_b=\textbf{I}_{r-r^\prime}$, $[\P^{(k)}_b \textbf{ 1}]^\top\hat{\P}^{(k)}_b=0$, and $(\hat{\Q}^{(k)}_b)^\top\hat{\Q}_b=\textbf{I}_{r-r^\prime}$, $[\Q^{(k)}_b \textbf{ 1}]^\top\hat{\Q}^{(k)}_b=0$. Now we obtain a closed-form update rule for $\X^{(k)}$:
\begin{align}
\label{x_update}
\X^{(k)} \leftarrow \sqrt{m}[\P^{(k)}_b,\text{ }\hat{\P}^{(k)}_b][\Q^{(k)}_b,\text{ }\hat{\Q}^{(k)}_b]^\top.
\end{align}
In practice, to compute such an optimal $\X^{(k)}$, we perform the eigendecomposition over the small
$r \times r$ matrix
$$(\bar{\B}^{(k)})^\top\bar{\B}^{(k)}=[\Q^{(k)}_b\text{ }\hat{\Q}^{(k)}_b]\begin{bmatrix}(\bSigma^{(k)})^2&\textbf{0}\\\textbf{0}&\textbf{0}\\\end{bmatrix}[\Q^{(k)}_b\text{ }\hat{\Q}^{(k)}_b]^\top,$$
which provides $\Q^{(k)}_b, \hat{\Q}^{(k)}_b, \bSigma^{(k)}$, and we can obtain $$\P_b^{(k)}=\bar{\B}^{(k)}\Q^{(k)}_b(\bSigma^{(k)})^{-1}.$$ Then matrix $\hat{\P}^{(k)}_b$ can be obtained by the aforementioned Gram-Schmidt orthogonalization. Note that it requires $O\big(r^2m\big)$ to perform SVD, Gram-Schimdt orthogonalization and matrix multiplication.

When $\D^{(k)}$ fixed, learning $\Y^{(k)}$ could be solved in a similar way:
\begin{align}
\max_{\Y^{(k)}} tr(\D^{(k)}(\Y^{(k)})^\top), \quad \textbf{1}_n^\top\Y^{(k)}=0, (\Y^{(k)})^\top\Y^{(k)}=n\I_r. \nonumber
\end{align}
We can obtain an analytic solution:
\begin{align}
\label{y_update}
\Y^{(k)} \leftarrow \sqrt{n}[\P^{(k)}_d,\text{ }\hat{\P}^{(k)}_d][\Q^{(k)}_d,\text{ }\hat{\Q}^{(k)}_d]^\top.
\end{align}
where each column of $\P^{(k)}_d$ and $\Q^{(k)}_d$ is the left and right singular vectors of $\bar{\D}^{(k)}$ respectively. $\hat{\Q}^{(k)}_d$ are the left singular vectors corresponding to zero singular values of the $r\times r$ matrix $(\bar{\D}^{(k)})^\top\bar{\D}^{(k)}$, and $\hat{\P}^{(k)}_d$ are the vectors obtained via the Gram-Schimidt process. We summarize the solution for CCCF in Algorithm \ref{cccf_al}.

\begin{algorithm}
\caption{The proposed algorithm for Compositional Coding for Collaborative Filtering (CCCF) .}
\label{cccf_al}
\begin{algorithmic}
\STATE{\textbf{Input:} $\R \in \mathbb{R}^{m\times n}$}
\STATE{\textbf{Output:} $\B^{(k)} \in\{\pm 1\}^{r\times m}, \D^{(k)} \in\{\pm 1\}^{r\times n}$}
\STATE{\textbf{Parameters:} number of components $G$, code length $r$, regularization coefficient $\alpha_1,\alpha_2$, bandwidth parameter $h$ }
\STATE{Initialize $\B^{(k)},\D^{(k)}$ and $\X^{(k)},\Y^{(k)}\in \mathbb{R}^{m\times n}$ by Eq. (\ref{init}).}
\WHILE{not converged}
\FOR{$k=1,\cdots,G$ \textbf{parallel}}
\STATE{Pick anchor points $(i^\prime_k,j^\prime_k)$.}
\FOR {$u=1,\cdots,m$}
\STATE{Update $\b_i^{(k)}$ according to (\ref{b_square})}
\ENDFOR
\FOR {$i=1,\cdots,n$}
\STATE{Update $\d_j^{(k)}$ according to (\ref{d_square}).}
\ENDFOR
\STATE{Update $\X^{(k)}$ and $\Y^{(k)}$ according to (\ref{x_update}) and (\ref{y_update}).}
\ENDFOR
\ENDWHILE
\end{algorithmic}
\end{algorithm}

\subsection{Initialization}
Note that the proposed optimization problem involves a mixed-integer non-convex problem, the initialization of model parameters plays an important role for fast convergence and for finding better local optimum solution. To achieve a good initialization in an efficient way, we essentially relax the binary constraints in Eq. (\ref{problem3}) into the following optimization: 

\begin{small}
\begin{align}
\label{init}
&\min\limits_{\U^{(k)},\V^{(k)},\X^{(k)},\Y^{(k)}}\;\sum_{(i,j)\in\mathcal{V}}\Big(R_{ij}-\sum^G_{k=1}w_{ij}^{(k)}(\u_i^{(k)})^\top\v_j^{(k)}\Big)^2\\
-2\alpha_1&\text{tr}((\B^{(k)})^\top\X^{(k)})-2\alpha_2\text{tr}((\V^{(k)})^\top\Y^{(k)})+\alpha_3\|\U^{(k)}\|^2_F+\alpha_4\|\V^{(k)}\|^2_F\nonumber\\
\text{s.t.}\; & \textbf{1}_m\X^{(k)}=0,\textbf{1}_n\X^{(k)}=0,(\Y^{(k)})^\top\Y^{(k)}=m\I_r,(\Y^{(k)})^\top\Y^{(k)}=n\I_r,\nonumber
\end{align}
\end{small}

We first initialize real-valued matrix $\U^{(k)}$ and $\V^{(k)}$ randomly and find the feasible solution for $\X^{(k)}$ and $\Y^{(k)}$ according to the above learning method with respect to $\X^{(k)}$ and $\Y^{(k)}$. Then the alternating optimization are conducted by updating $\U$ and $\V$ with traditional gradient descent method and updating $\X$ and $\Y$ with respect to the similar learning method. Once we obtain the solution $(\U^{(k)}_0,\V^{(k)}_0,\X^{(k)}_0,\Y^{(k)}_0)$, we can initialize CCCF with respect to component $k$ as:
\begin{small}
\begin{align}
\label{init1}
\B^{(k)} \leftarrow \text{sgn}(\U^{(k)}_0),\D^{(k)} \leftarrow \text{sgn}(\V^{(k)}_0),\X^{(k)} \leftarrow \X^{(k)}_0,\Y^{(k)} \leftarrow \Y^{(k)}_0.
\end{align}
\end{small}
The effectiveness of the proposed initialization will be discussed in Section \ref{settings} (illustrated in Figure \ref{init1}).

\vspace{-2pt}
\subsection{Distance Function}
\label{distance}
Previously we assume a general distance function $d$, which is defined to measure the distance between users or items so as to compute the component weights $w_i^{(k)}$ and $v_j^{(k)}$ in Eq. (\ref{weight}). The metric can be constructed with
side information, like users' social link \cite{zhao2017collaborative,wang2016social} or using metric learning techniques \cite{xing2003distance}. However, many datasets do not include such data. In this work, we follow the idea of \cite{lee2013local,lee2014local} which factorizes the observed interaction matrix using MF and obtain two latent representation matrices $\U$ and $\V$ for users and items, respectively. Then the distance between two users can be computed by the cosine distance between the obtained latent representations, which is formulated as $
d(u_i,u_j)=arccos\Big(\frac{\langle\u_i,\u_j\rangle}{\|\u_i\|\cdot\|\u_j\|}\Big) 
$. The distance between two items can be computed in the same way.

\vspace{-2pt}
\subsection{Complexity}
The computational complexity of training CCCF is $K$ times the complexity of learning each set of binary codes. It converges quickly in practice, which usually takes about $4\sim5$ iterations in our experiments. For each iteration, the computational cost for updating $\B^{(k)}$ and $\D^{(k)}$ is $O\big(\#iter (m+n)r^2\big)$. In practice, $\#iter $ is usually $2\sim5$. The computational cost for updating $\X^{(k)}$ and $\Y^{(k)}$ is $O\big(r^2m\big)$ and $O\big(r^2n)$, respectively. Suppose the entire algorithm requires $T$ iterations for convergence, the overall time complexity for Algorithm \ref{cccf_al} is $O(Tqr^2(m+n))$, where we found $T$ empirically is no more than $5$. In summary, CCCF is efficient and scalable because it scales linearly with the number of users and items.
\vspace{-2pt}
\subsection{Fast Retrieval via Integer Weight Scaling}
\label{si}
Floating-point operations over user and item weight vectors invoke more CPU cycles and are usually much slower than integer computation. The cost of top-$k$ recommendation would be remarkably lower when the scalars in the weight vectors are integers instead of floating numbers. An intuitive way is to adopt integer approximation via rounding the scalars in user and item weight vectors.  However, if the weights are too small, it will incur large deviation. To tackle this problem, we scale the original scalars by multiplying each weight by $e$ and then approximate them with integers in  preprocessing,
\begin{small}
\begin{align}
\hat{\bbeta}_i=\Big\{\nint{e\cdot \eta^{(1)}_i},\dots,\nint{e\cdot \eta^{(G)}_i}\Big\},\quad \hat{\bxi}_j=\Big\{\nint{e\cdot \xi^{(1)}_j}, \dots, \nint{ e\cdot \xi^{(G)}_j}\Big\}, \nonumber
\end{align}
\end{small}
where $\nint{e\cdot \eta^{(k)}_i}$ is a round function to obtain an integer approximation with respect to $e\cdot \eta^{(k)}_i$.


\section{Experiments}
In order to validate the effectiveness and efficacy of the proposed CCCF method for recommender systems, we conduct an extensive set of experiments to examine different aspects of our method in comparison to state-of-the-art methods based on conventional binary coding. We aim to answer the following questions:
\begin{itemize}
\item[\textbf{RQ1:}] How does CCCF perform as compared to other state-of-the-arts hashing based recommendation methods in terms of both accuracy and retrieval time?

\item[\textbf{RQ2:}] How do different hyper-parameter settings  (e.g., number of components,
 and code length) affect the accuracy of CCCF ?

\item[\textbf{RQ3:}] How do the sparsity of component weight vectors (controlled by the bandwidth parameter $h$) and integer scaling (controlled by parameter $e$) affect both the accuracy and retrieval cost of CCCF? How to choose optimal values ?

\item[\textbf{RQ4:}]  Does the representation of compositional codes in CCCF enjoy a much stronger representation capability than the traditional binary codes in DCF given the same model size ?

\end{itemize}

\subsection{Experimental Settings}
\label{settings}

\subsubsection{\textbf{Datasets and Settings.}}
We run our experiments on three public datasets: Movielens 1M\footnote{http://grouplens.org/datasets/movielens}, Amazon and Yelp\footnote{
http://www.yelp.com/dataset challenge} which are widely used in the literature. All of these ratings range from $0$ to $5$. Considering the severe sparsity of Yelp and Amazon original
datasets, we followed the conventional filtering strategy \cite{rendle2009bpr} by removing users and items with less than $10$ ratings.  The statistics of the filtered datasets are shown in Table \ref{tbl:dataset}. For each user, we randomly sampled $70\%$ ratings as training data and the rest $30\%$ for test. We repeated for $5$ random splits and reported the averaged results.

\begin{table}[htb]
\centering
\begin{small}
\vspace{-0.1cm}
\begin{tabular} {|l|c|c|c|c|}
\hline
Dataset & \#Ratings & \#Items & \#Users  & \#density \\
\hline
\hline
Movielens 1M    & 1,000,209    & 3900 	& 6040	  &4.2\%\\
Yelp    & 696,865    & 25,677 	& 25,815     &0.11\%\\
Amazon         &5,057,936       & 146,469      & 189,474       &0.02\%\\
\hline
\end{tabular}
\end{small}
\caption{Summary of datasets in our experiments.}
\label{tbl:dataset}
\end{table}
\subsubsection{\noindent\textbf{Parameter Settings and Performance Metrics.}}
For CCCF, we vary the number of components and the code length of each component in range $\{4,8,12,16\}$. The hyper-parameters $\alpha$ and $\beta$ are tuned within $\{10^{-4},10^{-3},\dots,10^{2}\}$. Grid search is performed to choose the best parameters on the training split. We evaluate our proposed algorithms
by Normalized Discounted Cumulative Gain (NDCG) \cite{jarvelin2000ir},
which is probably the most popular ranking metric for capturing the importance of retrieving good items at the top of ranked lists. The average NDCG at cut off $[2,4,6,8,10]$ over all users is the final metric. A higher NDCG@K reflects a better accuracy of recommendation performance.

\subsubsection{\noindent\textbf{Baseline Methods and Implementations.}}
To validate the effectiveness of CCCF, we compare it with several state-of-the-art real-valued CF methods and hashing-based CF methods:
\begin{itemize}[leftmargin=*]
\item\noindent\textbf{MF}: This is the classic Matrix Factorization based CF algorithm \cite{koren2009matrix}, which learns real-valued user and item latent vectors in Euclidean space.

\item\noindent\textbf{BCCF}: This is a two-stage binarized CF method \cite{zhou2012learning} with a relaxation stage and a quantization stage. At these two stages, it successively solves MF with balanced code regularization and applies orthogonal rotation to obtain user codes and item codes. 

\item\noindent\textbf{DCF}: This is the first method \cite{zhang2016discrete} directly tackles a discrete optimization problem for seeking informative and compact binary codes for users and items.

\item\noindent\textbf{DCMF}: This is the state-of-the-art binarized method \cite{lian2017discrete} for CF with side information. It extends DCF by encoding the side features as the constraints for user codes and item codes.
\end{itemize}

\begin{figure}[ht]
  \centering
  \hspace{-0.2cm}
    \includegraphics[width=0.21\textwidth]{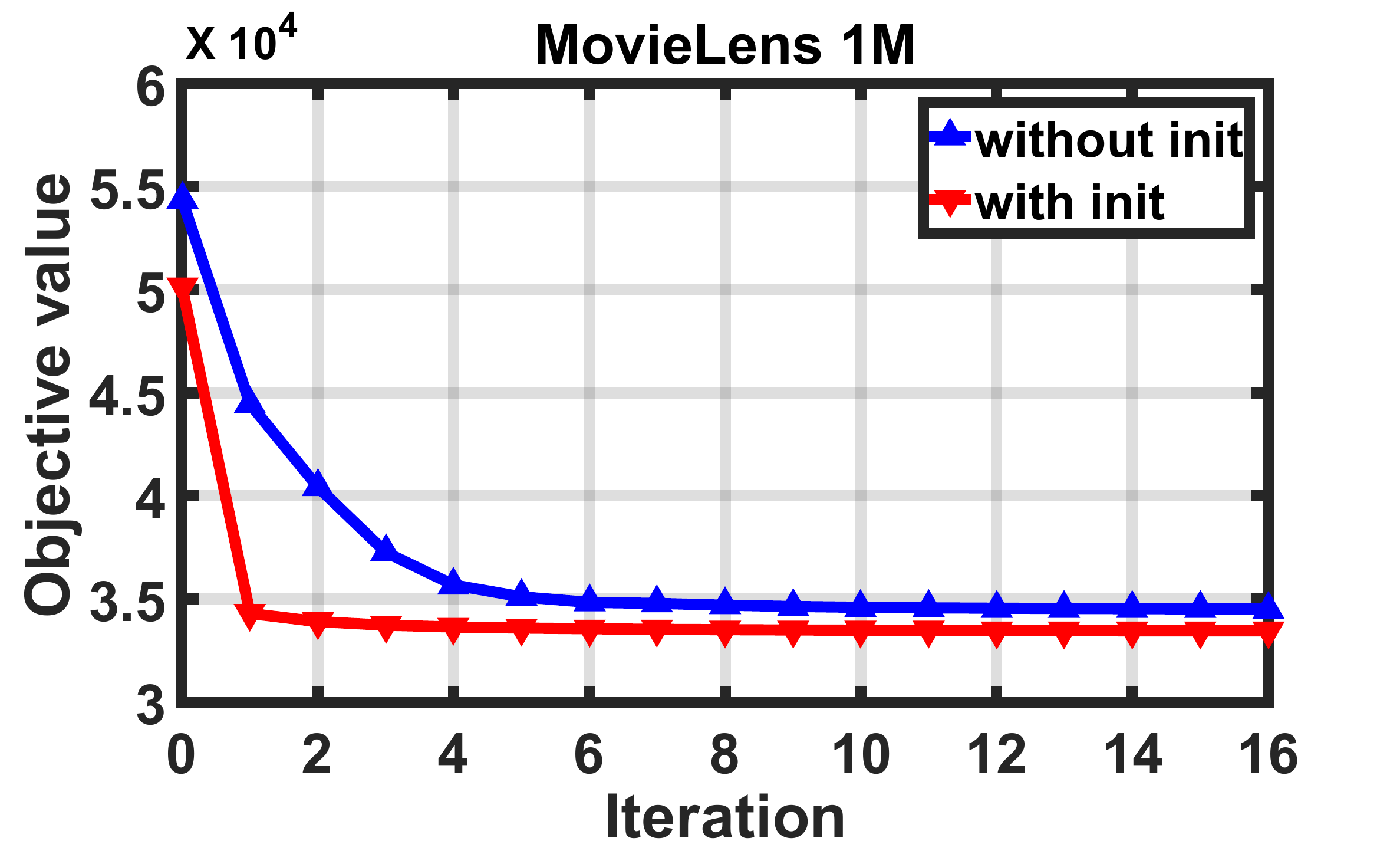}
    \hspace{-0.10cm}
    \includegraphics[width=0.21\textwidth]{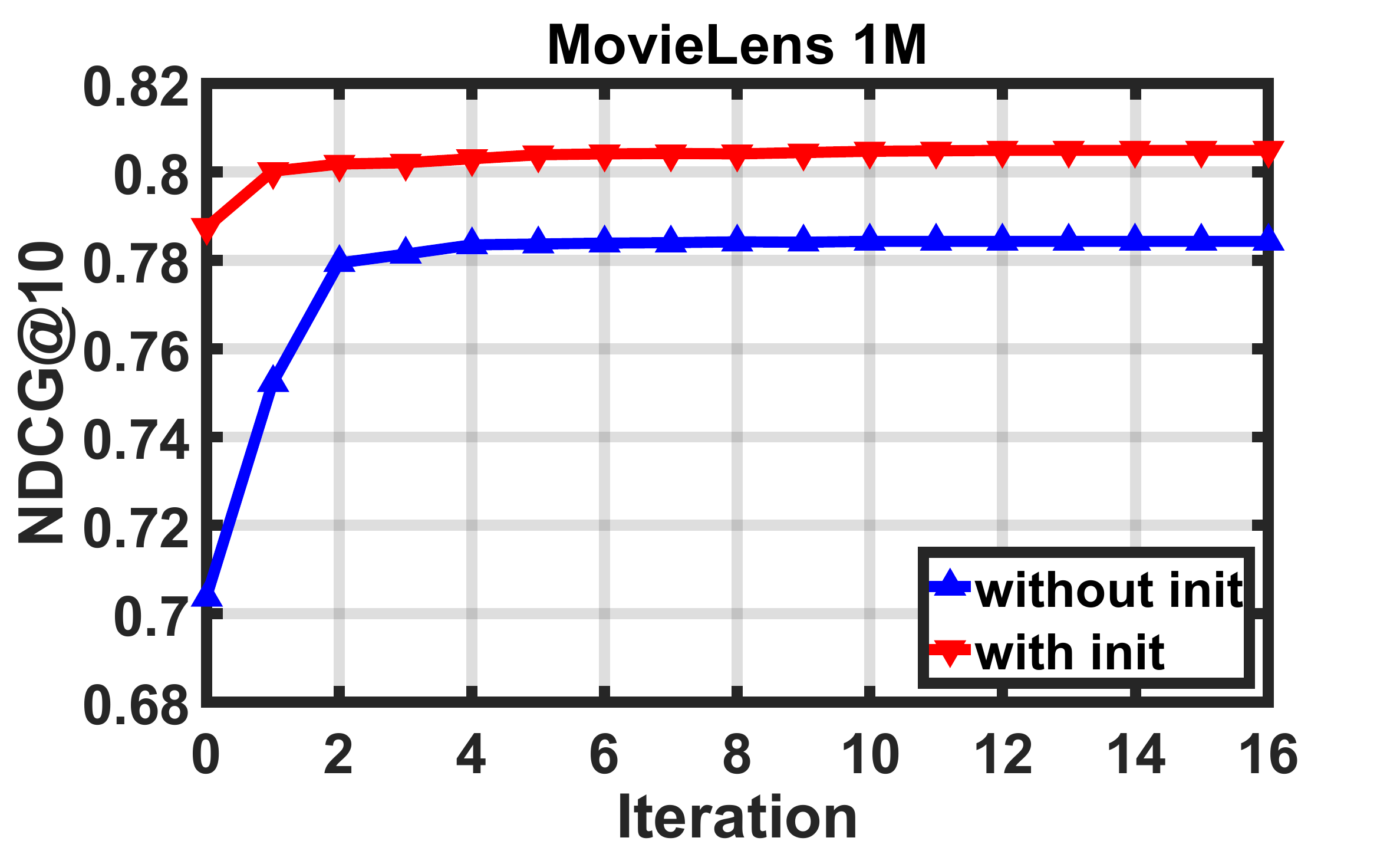}
  \caption{Convergence of the overall objective values and NDCG@10 of CCCF with/without initialization on the Movielens 1M dataset. The use of the proposed initialization leads to faster convergence and better results.
  }
\label{init2}
\end{figure}

The CCCF algorithm empirically converges very fast and using the initialization generally helps as shown in Figure \ref{init2}.

\subsection{Experimental Results}
\subsubsection{\bf Comparisons with State-of-the-arts (RQ1)}
\begin{figure*}[!htb]
  \centering
    \includegraphics[width=0.245\textwidth]{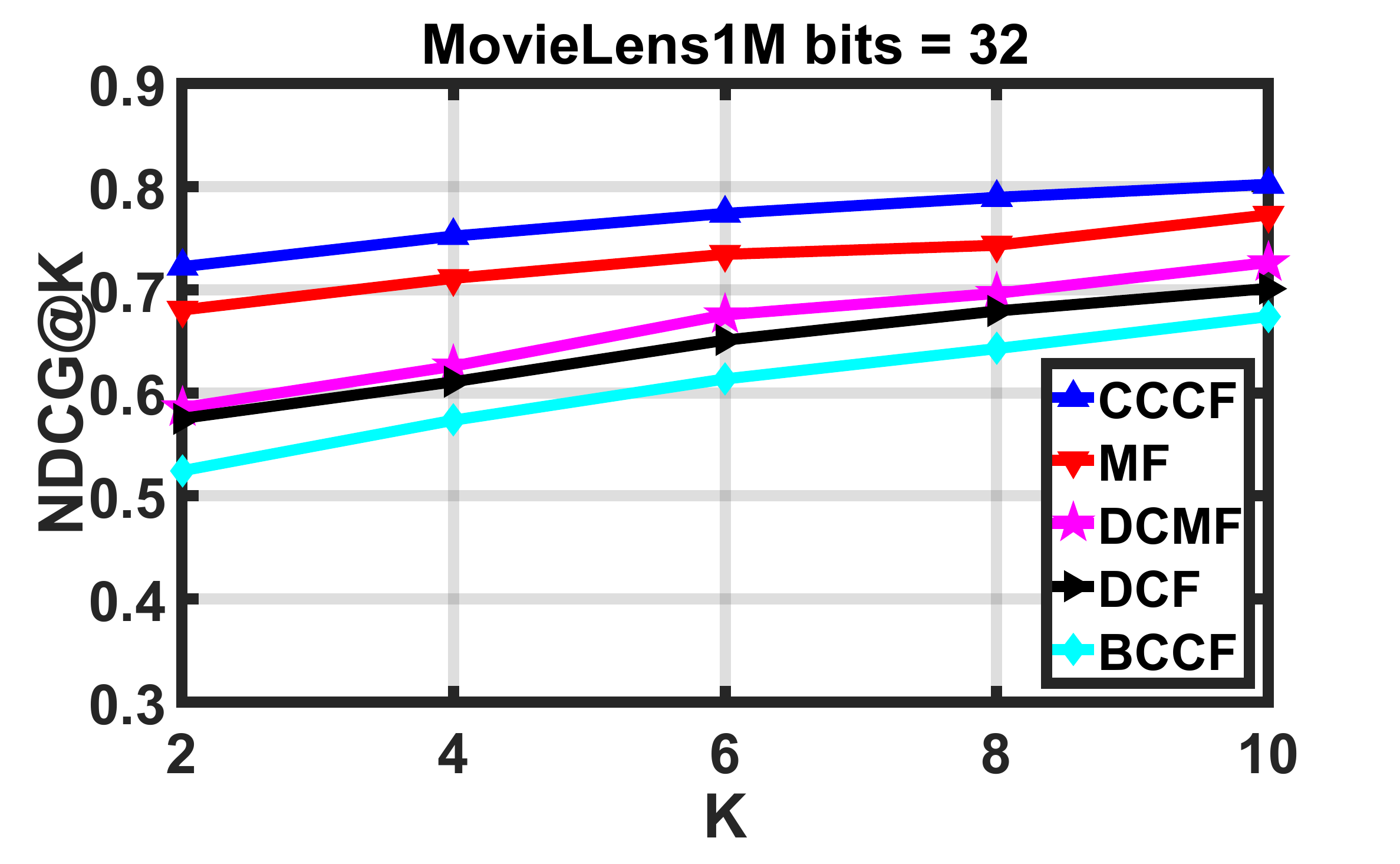}
    \includegraphics[width=0.245\textwidth]{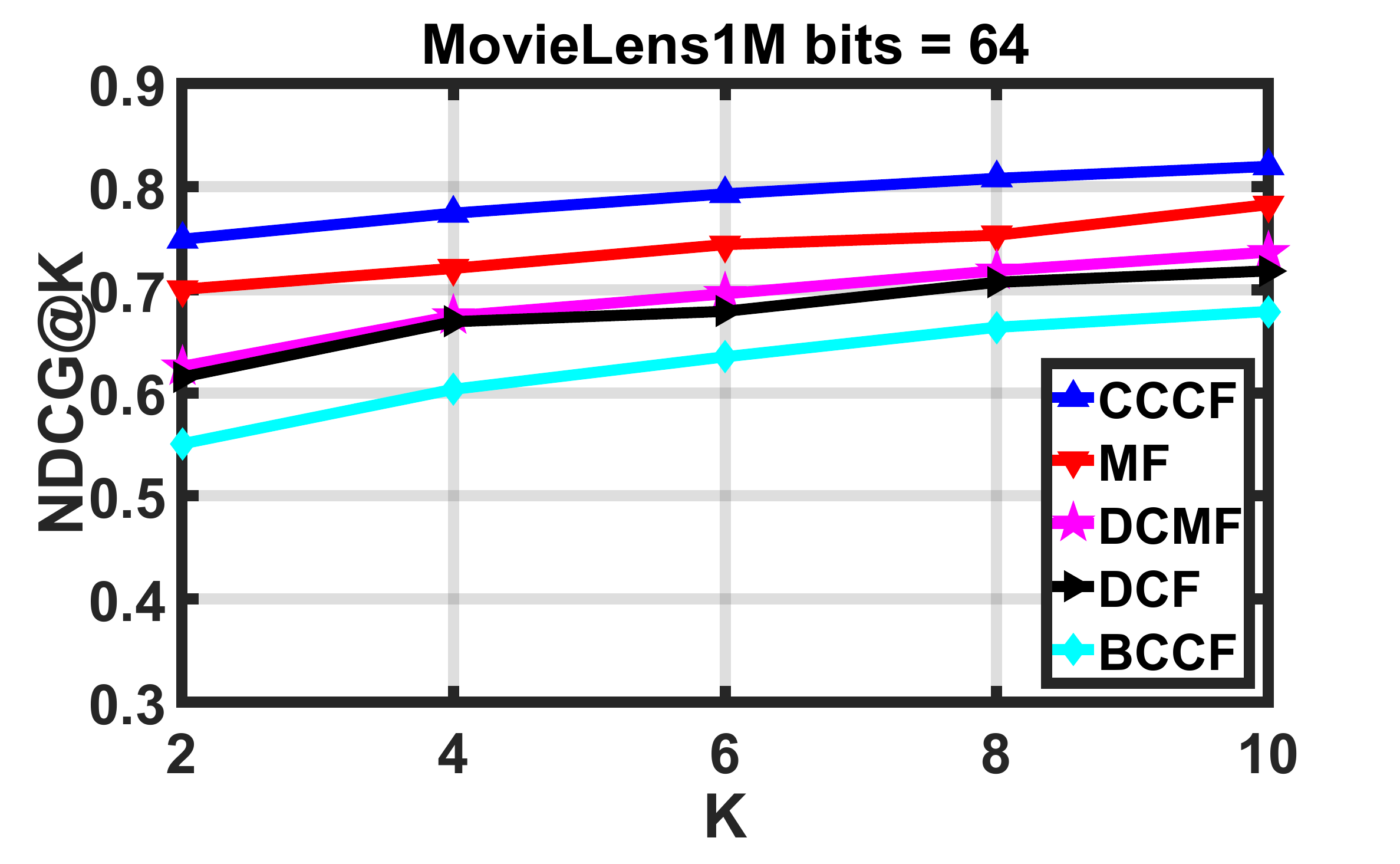}
    \includegraphics[width=0.245\textwidth]{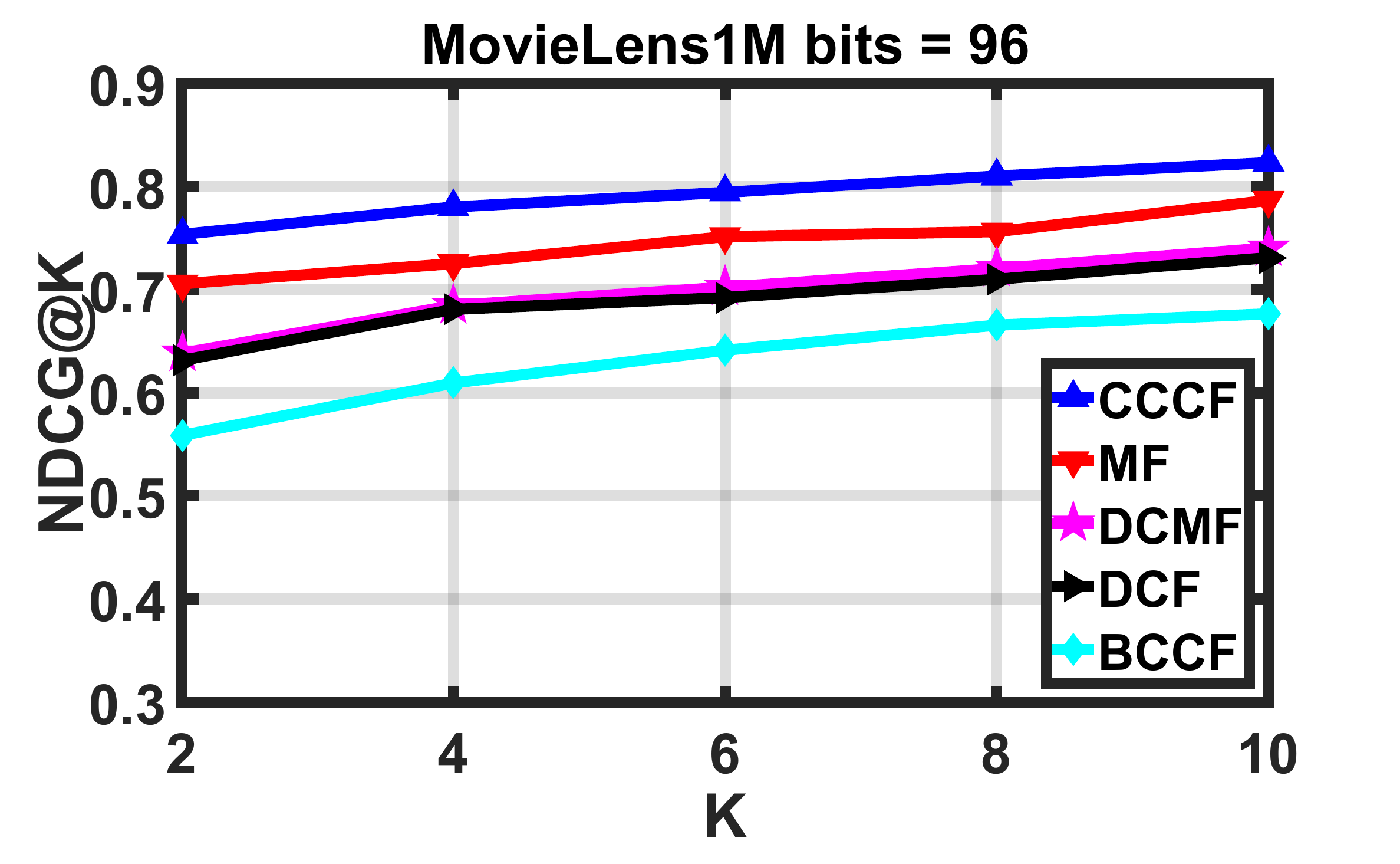}
    \includegraphics[width=0.245\textwidth]{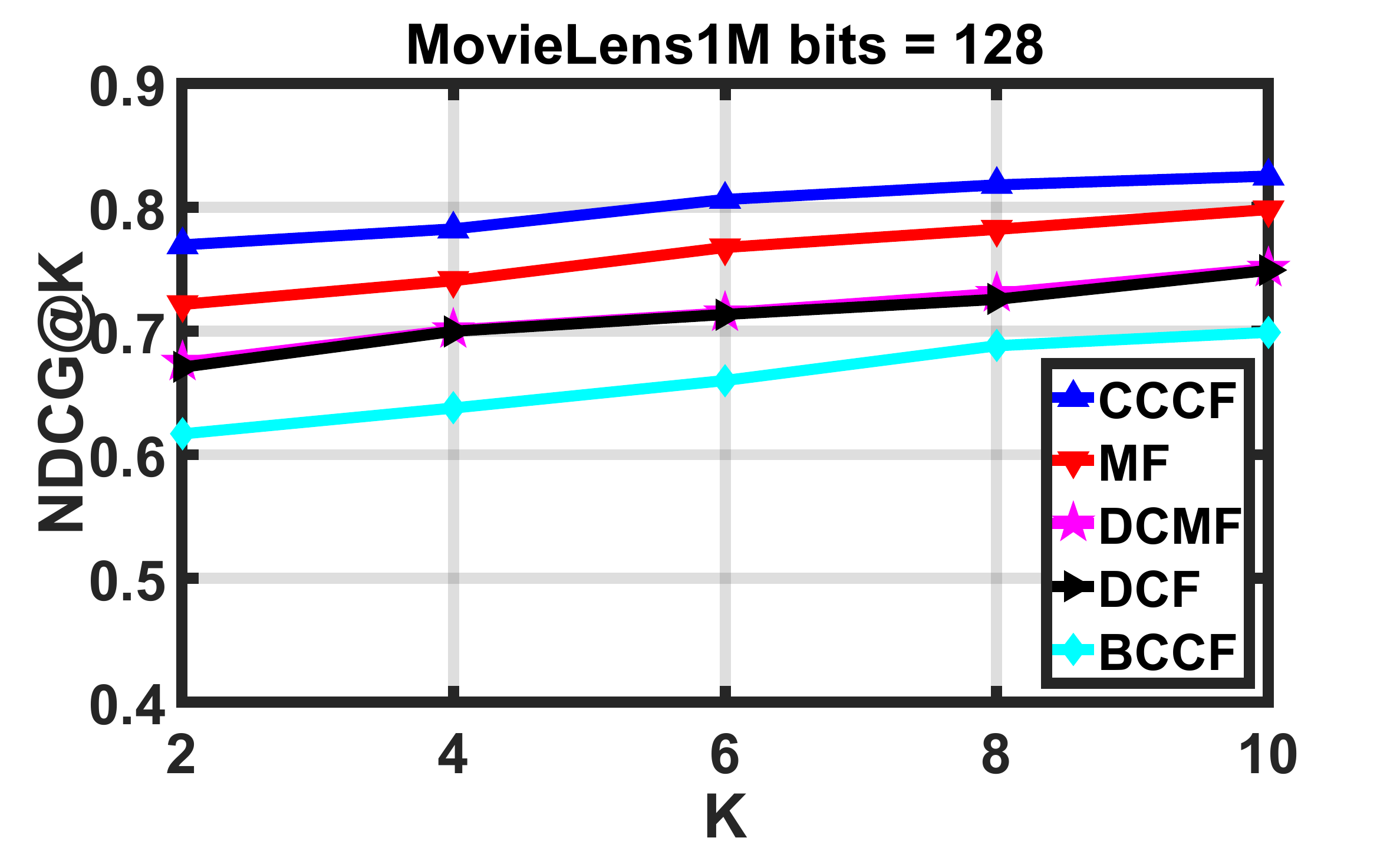}
    \includegraphics[width=0.245\textwidth]{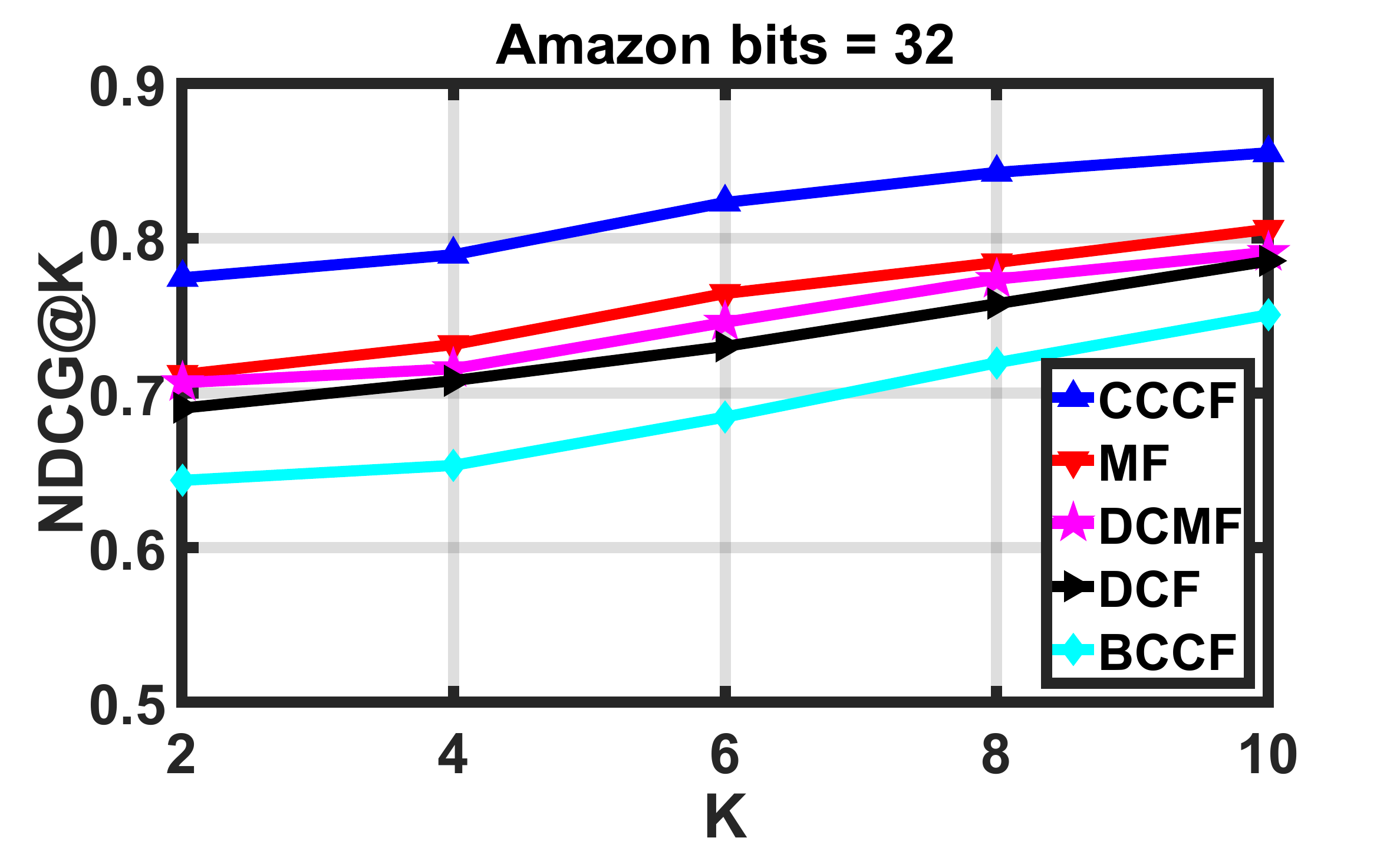}
    \includegraphics[width=0.245\textwidth]{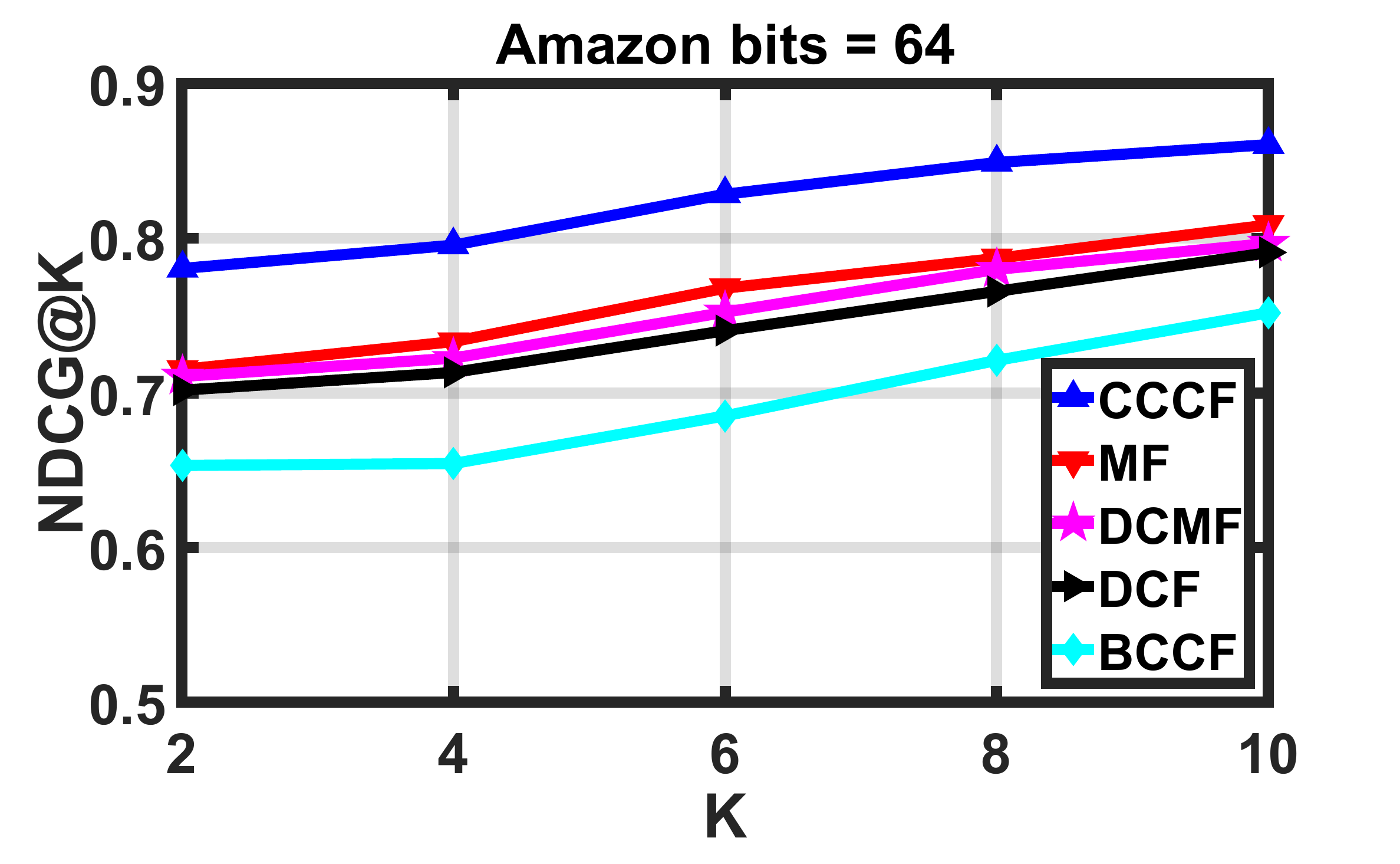}
    \includegraphics[width=0.245\textwidth]{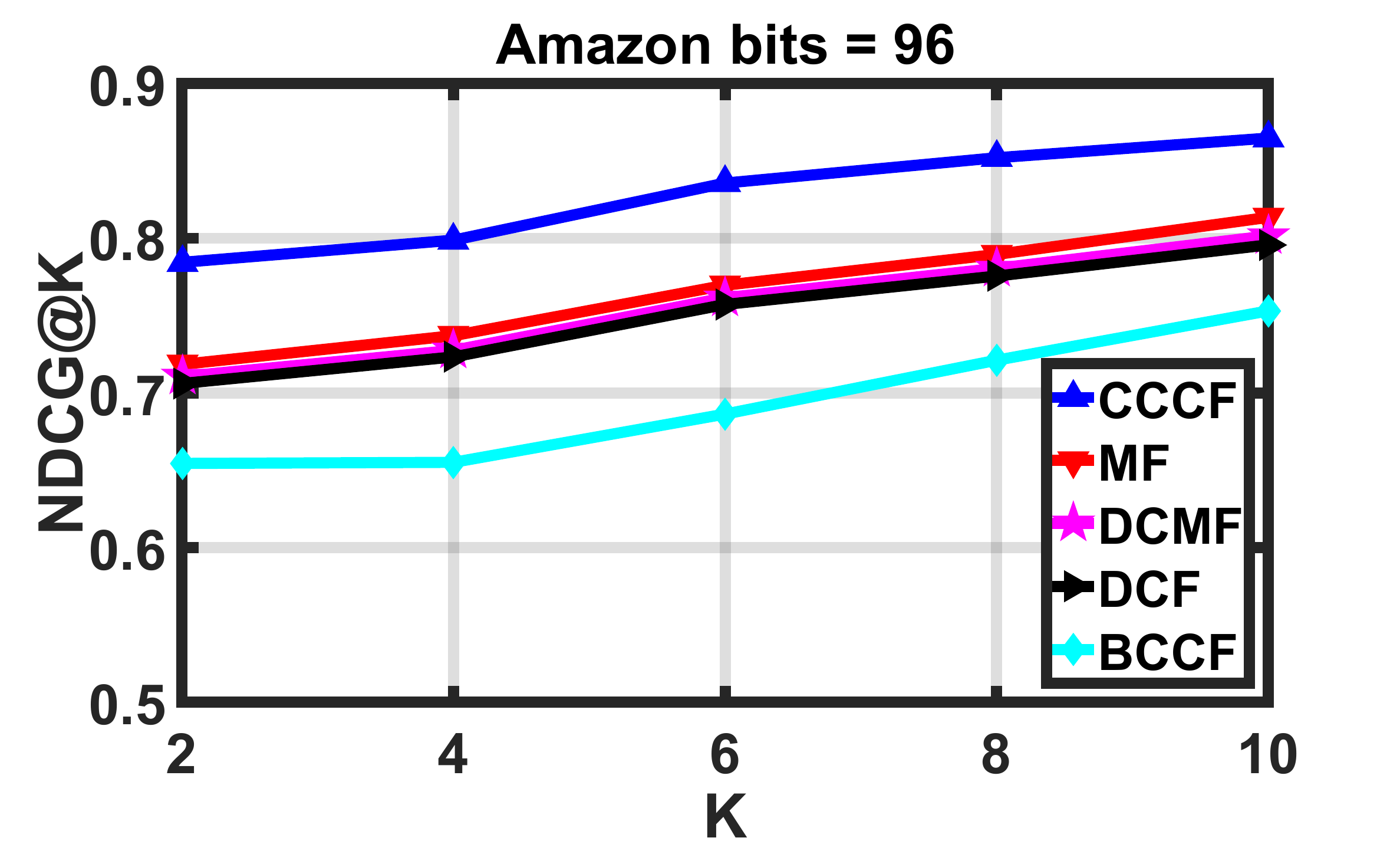}
    \includegraphics[width=0.245\textwidth]{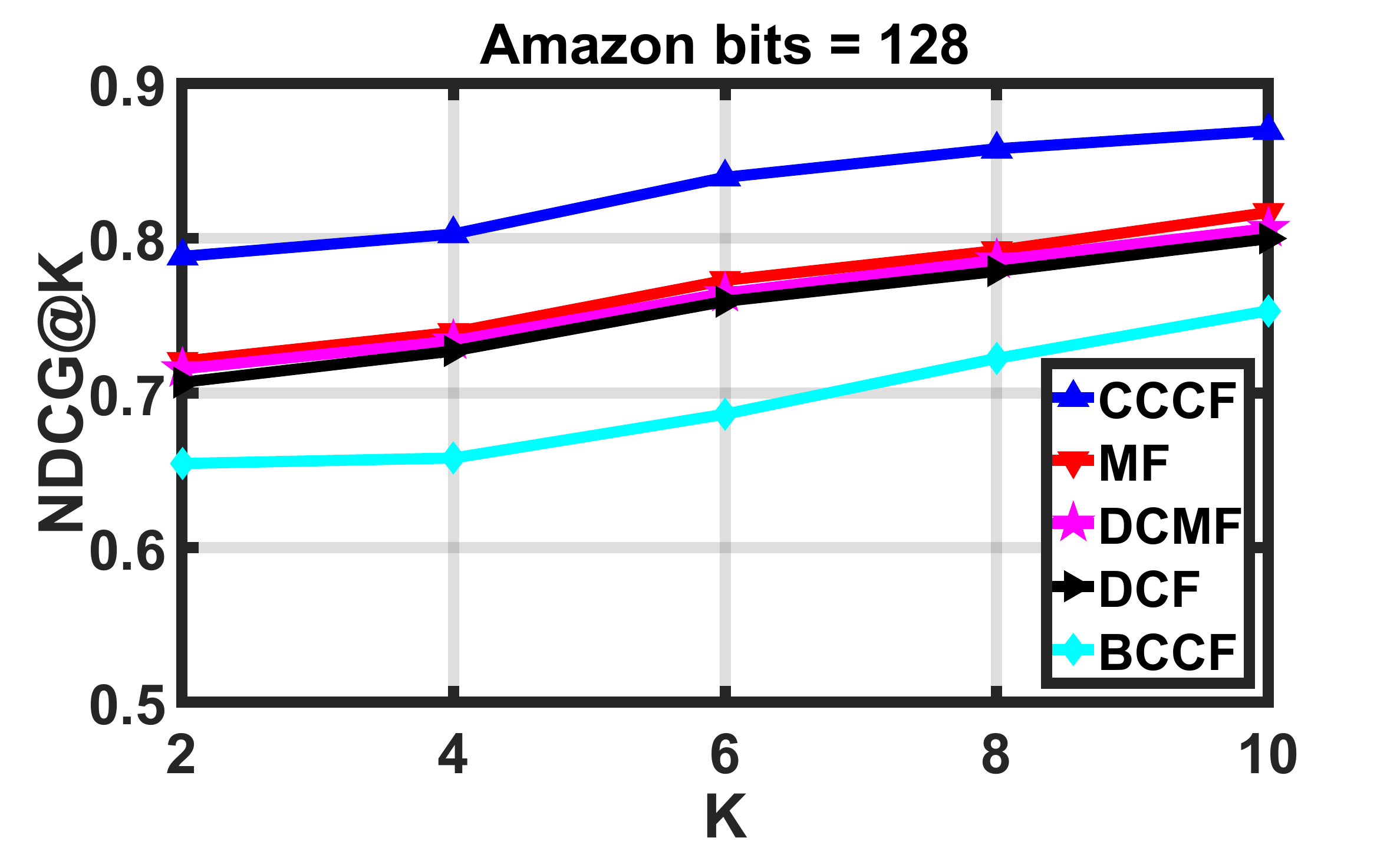}
    \includegraphics[width=0.245\textwidth]{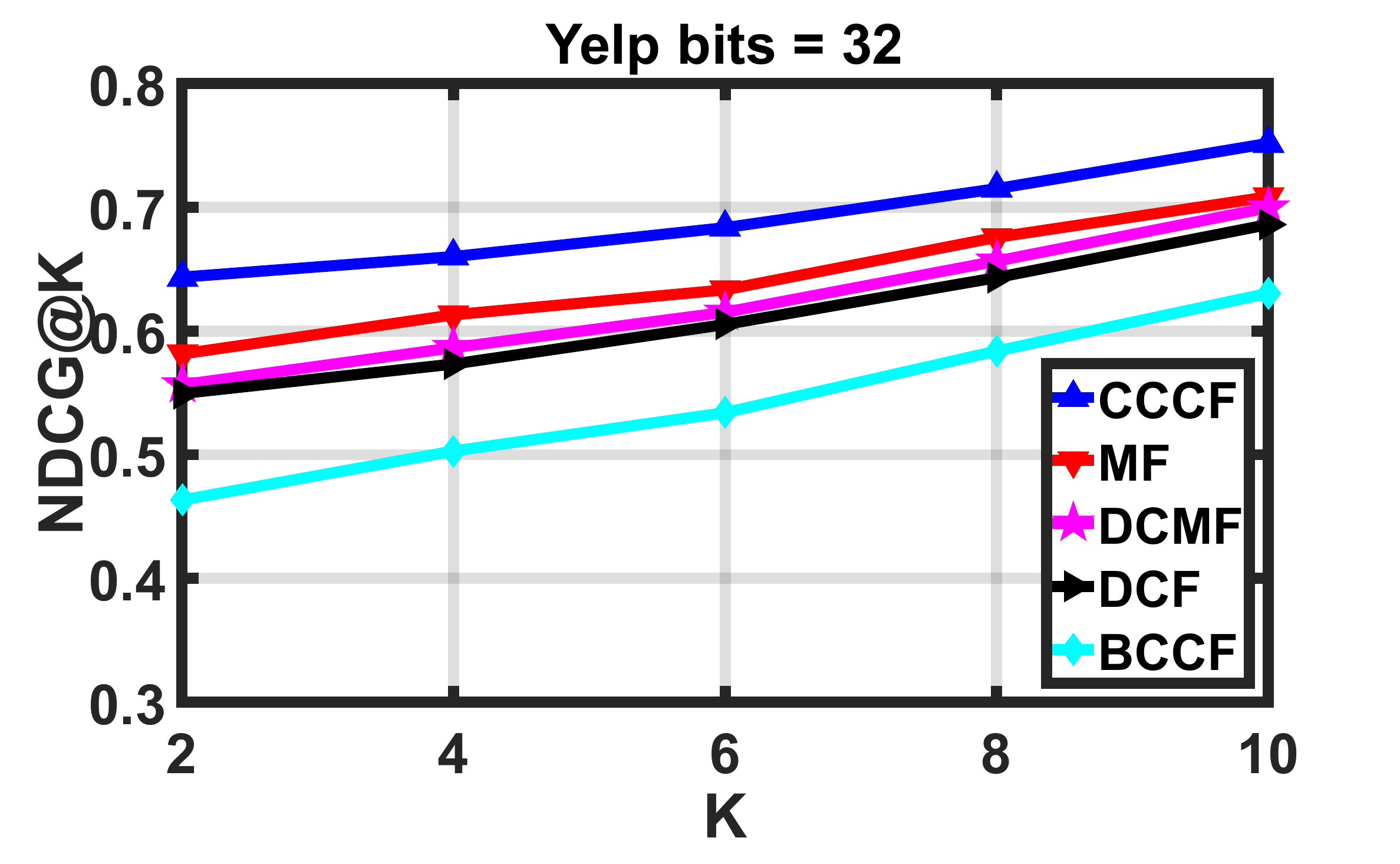}
    \includegraphics[width=0.245\textwidth]{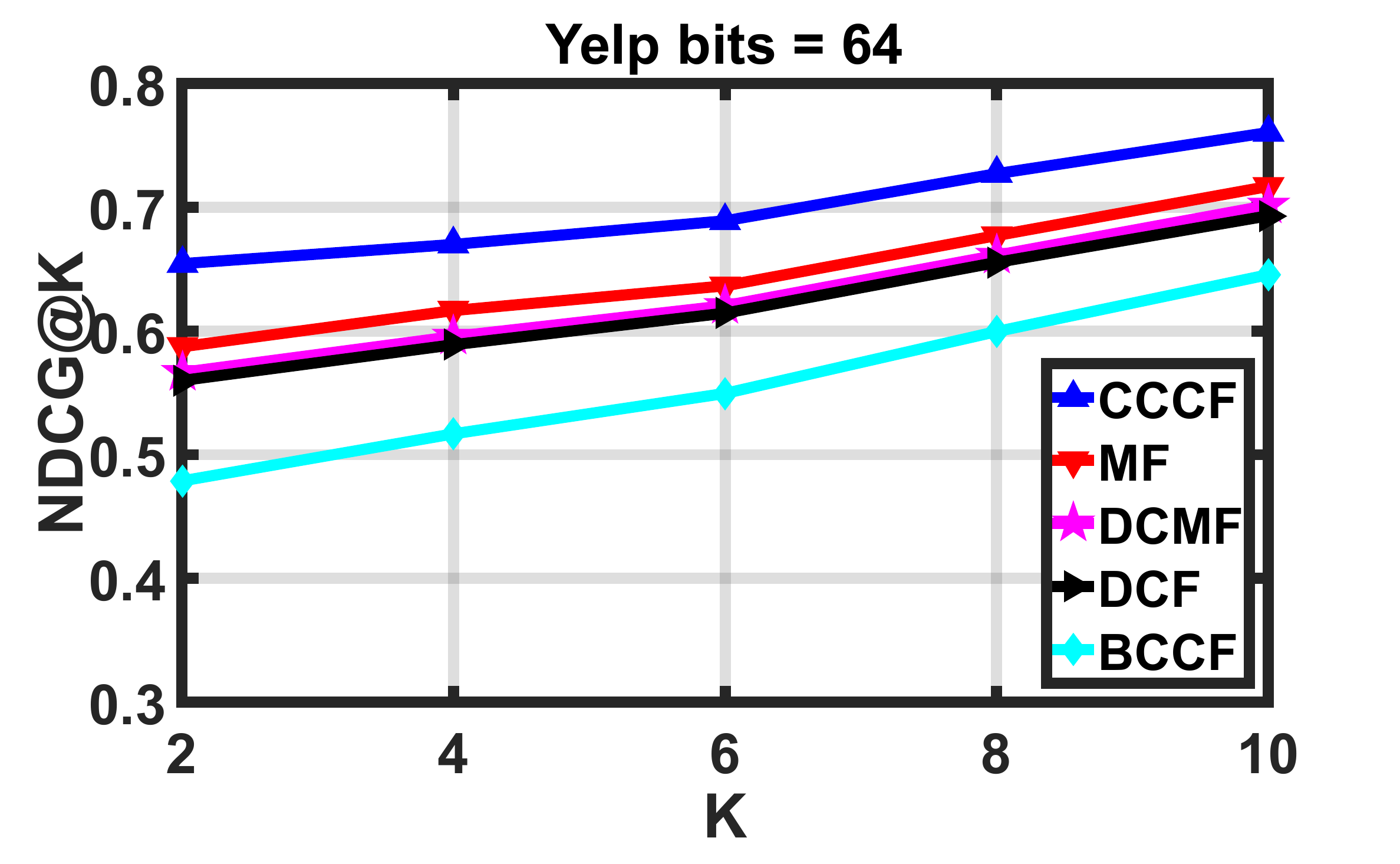}
    \includegraphics[width=0.245\textwidth]{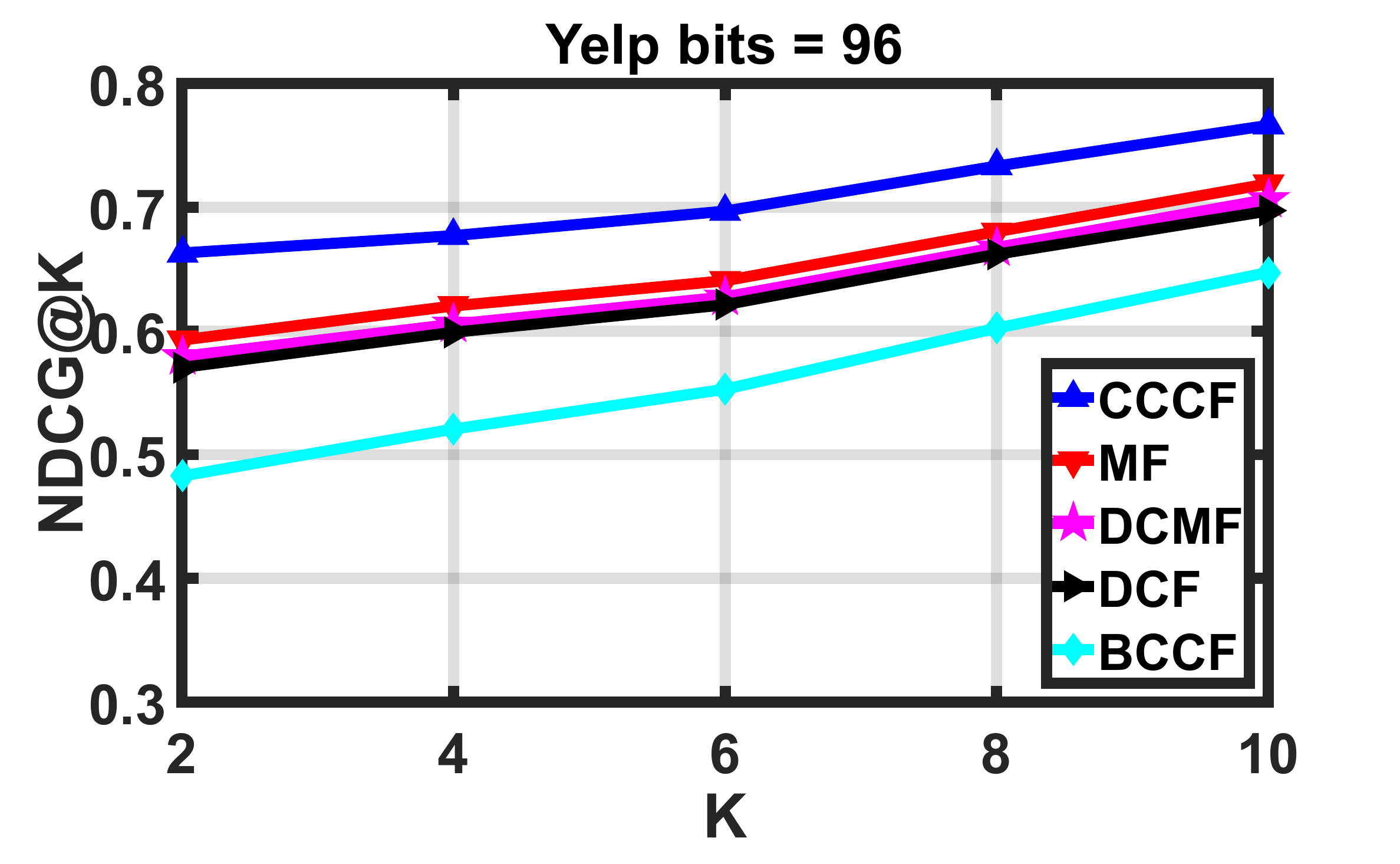}
    \includegraphics[width=0.245\textwidth]{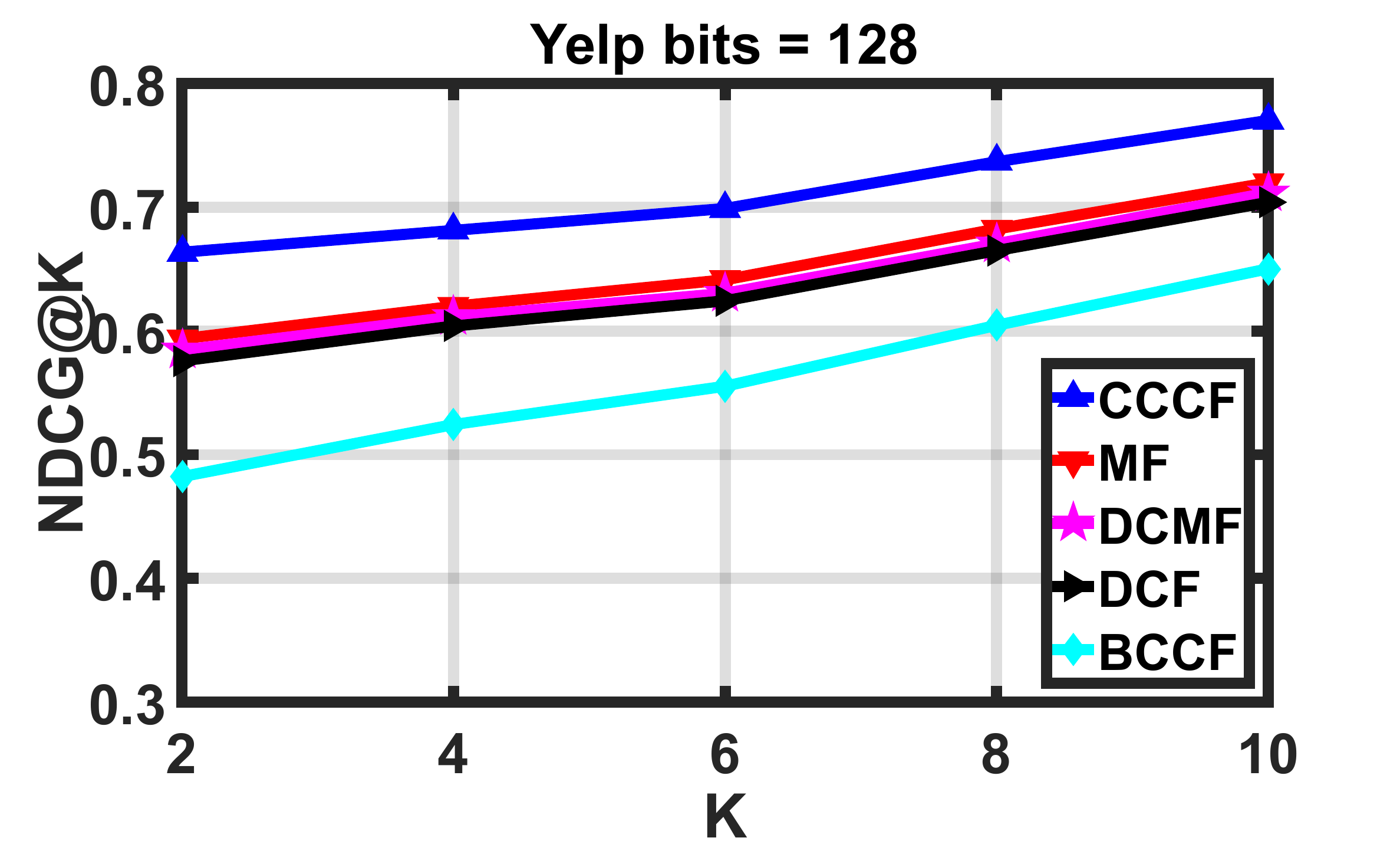}
  \caption{Item recommendation performance comparison of NDCG@K with respect to code length in bits.}
  \label{fig:pre}
\end{figure*}
\begin{figure*}[!htb]
  \centering
  \subfigure[\textbf{Varied total bits (rG)}]{
    \includegraphics[width=0.23\textwidth]{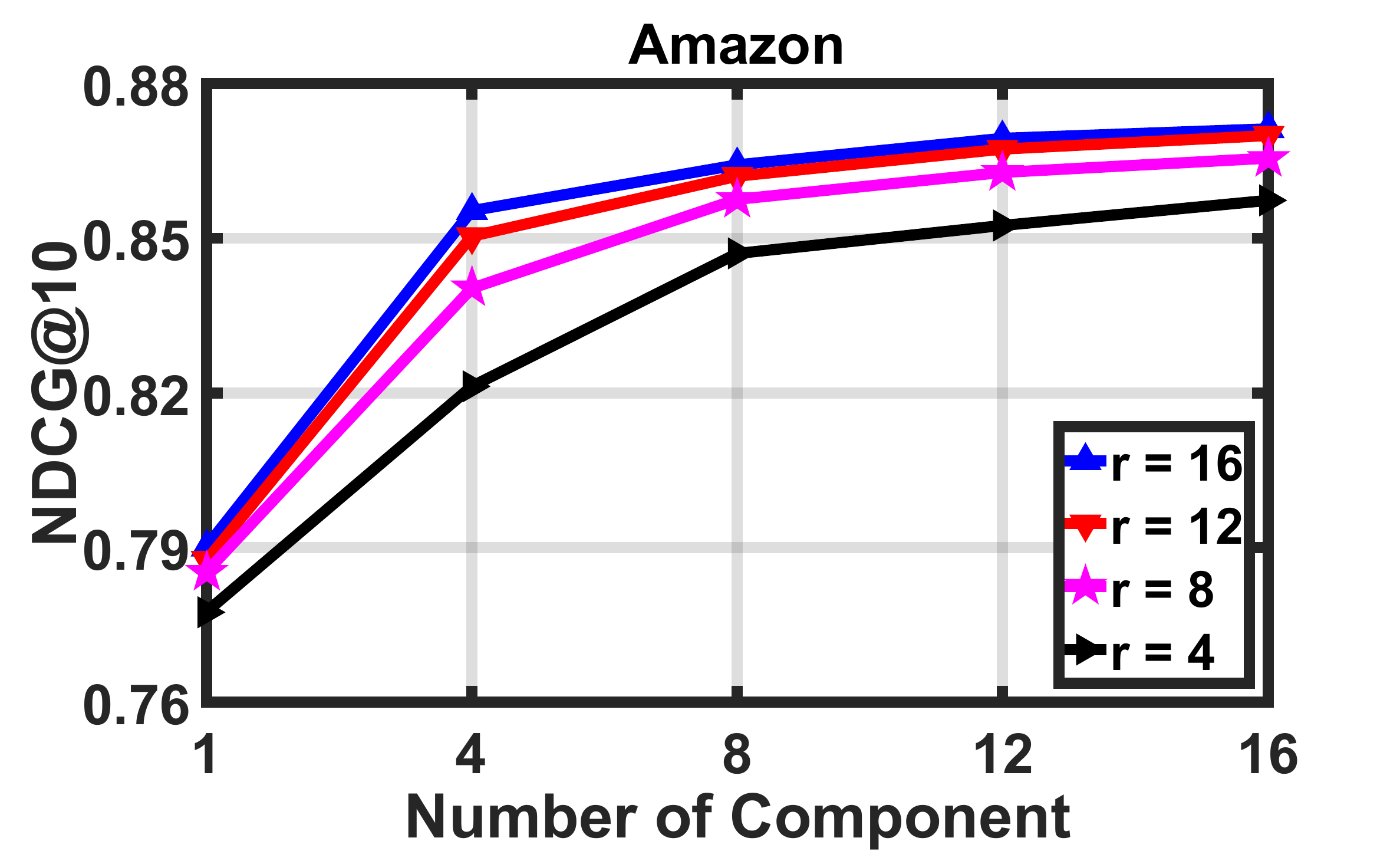}
    \includegraphics[width=0.23\textwidth]{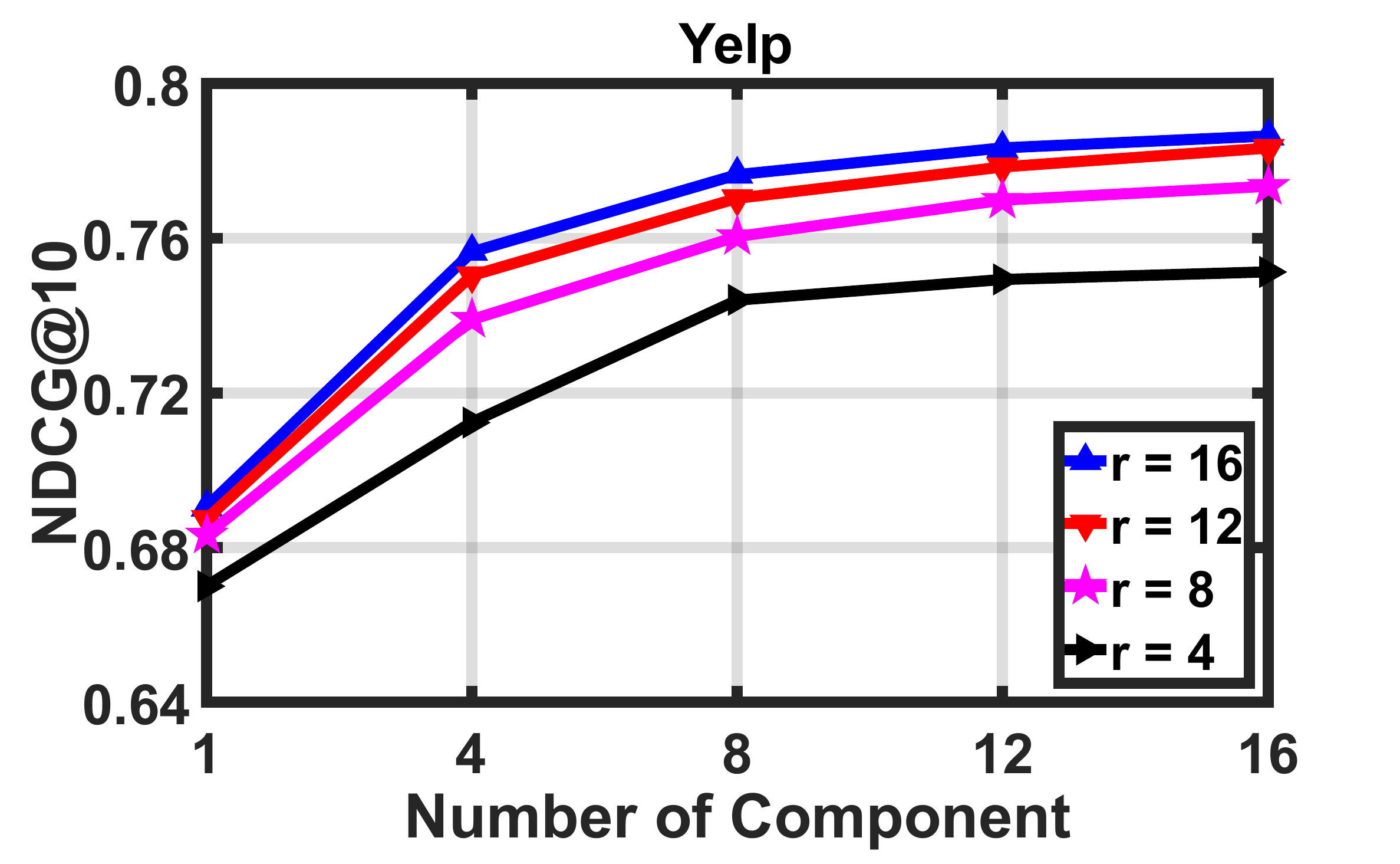}
    \label{hyper1}
    }
  \subfigure[\textbf{Fixed total bits (rG)}]{\label{hyper2}
    \includegraphics[width=0.23\textwidth]{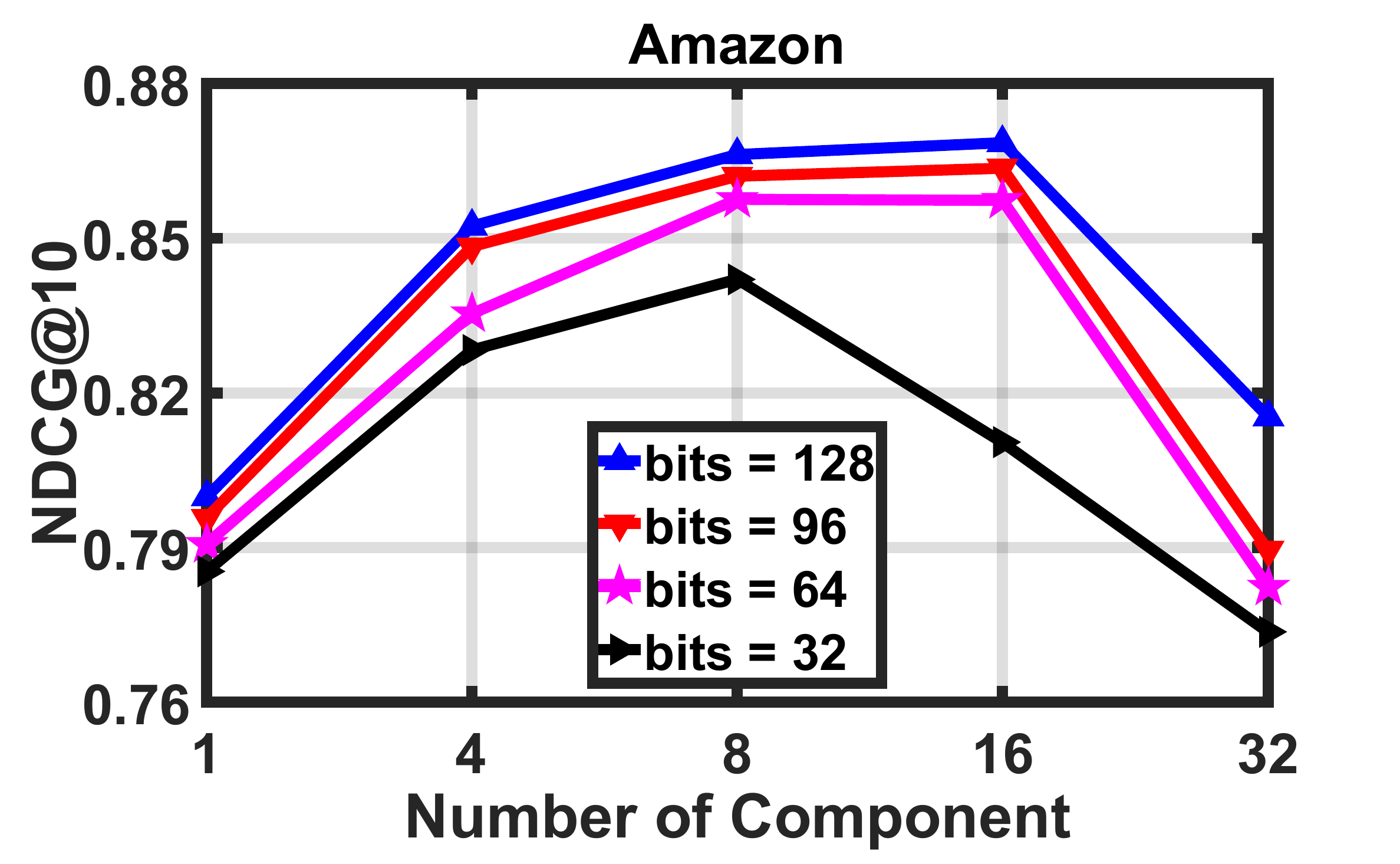}
    \includegraphics[width=0.23\textwidth]{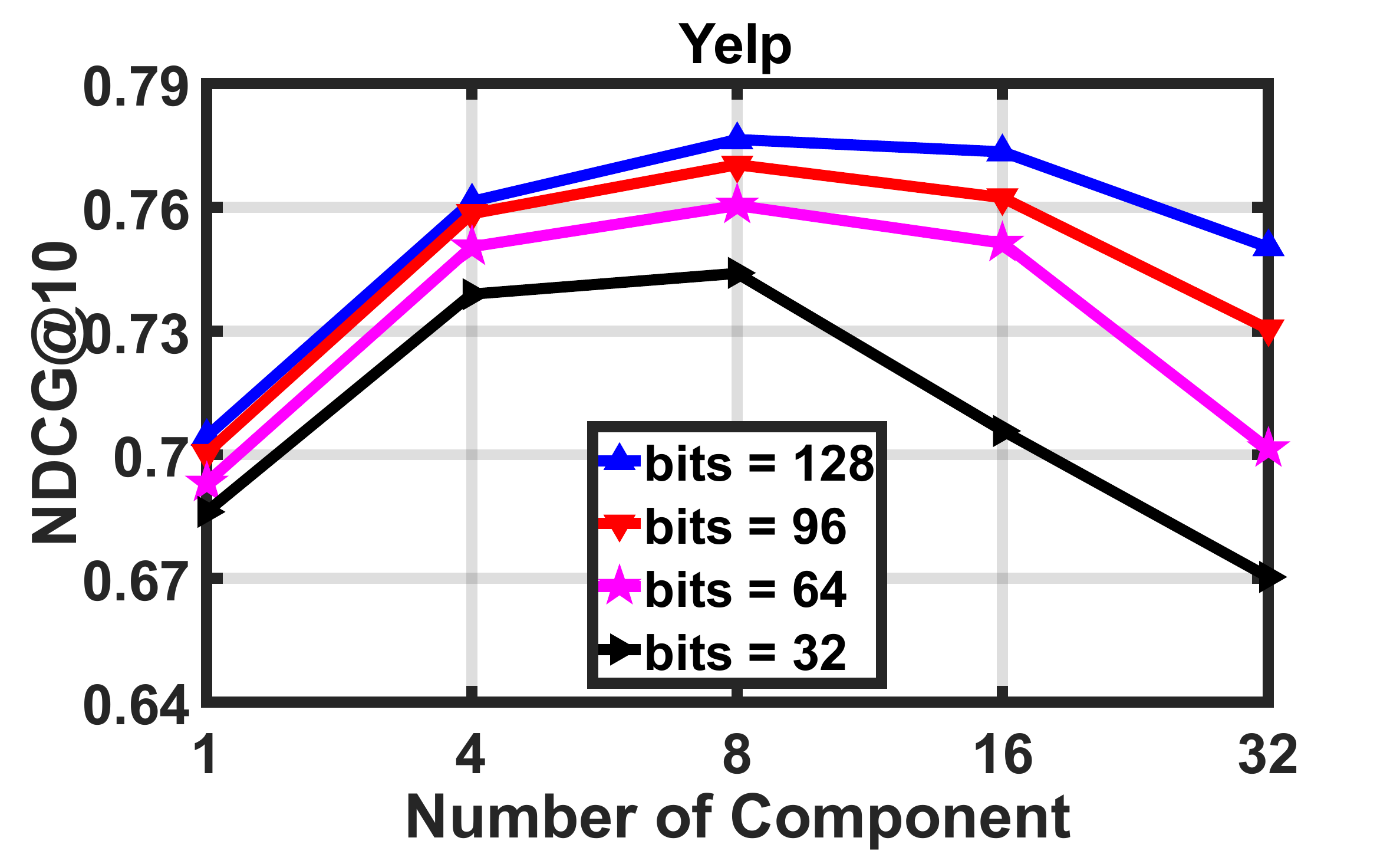}
  }
  \caption{Performance of CCCF with respect to code length in bits (r) and number of components (G).}
\label{hyper}
\end{figure*}
Figure \ref{fig:pre} shows the results of top-$k$ recommendation with $k$ setting from $2$ to $10$. Note that the number of components $G$ is fixed to $8$ and the code length of each component varies in $\{4,6,8,10,12,14,16\}$. For fair comparison,  the code length of DCF and the rank of MF are equal to the total bits of CCCF, which are equivalent to the summation of each component's code length of CCCF ($rG$), so that the performance gain is not from increasing model complexity. From Figure \ref{fig:pre},  we can draw the following major observations:

First of all, we observe that CCCF considerably outperforms BCCF, DCF and DCMF which are the state-of-the-art hashing based CF methods. The performance of CCCF and DCF  continiously increase as we increase the code length. Surprisingly, CCCF can even achieve remarkable superior performance using only $32$ bits in comparison to the DCF using $128$ bits on three datasets. This improvement indicates the impressive effectiveness of learning compositional codes.

Second, between baseline methods, DCF consistently outperforms
BCCF, while slightly underperforms DCMF. This is consistent with the findings in \cite{zhang2016discrete} that the performance of direct discrete optimization could surpass that of the two-stage methods. Besides, side information makes user codes and item codes more representative, which improves the recommendation performance.

Moreover, it is worth mentioning that CCCF outperforms MF, which is a real-valued CF method, particularly on the Amazon and Yelp dataset. The reasons for this are two-fold. First, compositional structure of CCCF has a much stronger representation capability which could discover complex relationships among users and items. Second, the higher sparsity of the dataset makes MF easy to overfit, whereras the binarized and sparse parameters in CCCF could alleviate this issue. This finding again demonstrates the effectiveness of our method.


\begin{table}[htb]
\centering
    \begin{tabular}{ l | l | l  l | l  l }
    \hline
    \multirow{2}{*}{Dataset} & CCCF &  \multicolumn{2}{c|}{MF}& \multicolumn{2}{c}{DCF}\\
    &\small{Time} &\small{Time}& \small{Speedup} &\small{Time}& \small{Speedup} \\
    \hline
    \hline
    \small{Movielens 1M} &7.12 & 49.66 & $\times$\small{6.97} & 10.64 & $\times$\small{1.49}\\
    \small{Amazon} &831.34 & 5917.56 & $\times$\small{7.12}& 1231.35 & $\times$\small{1.48}\\
    \small{Yelp} &184.48&  1450.25 & $\times$\small{7.86}& 264.65 & $\times$\small{1.43} \\
    \hline
    \end{tabular}
    \caption{Retrieval time (in seconds) of recommendation methods on three datasets, 'Speedup' indicates the speedup ($\times$) of CCCF ($G=8, r=16$) over baselines.}
    \label{retrieval}
    \vspace{-0.1cm}
\end{table}
\vspace{-0.3cm} 
Finally, Table \ref{retrieval} shows the total time cost (in seconds) taken by each method to generate the top-$k$ item list of all items. Overall, the hashing based methods (DCF and CCCF) outperform real-valued methods (MF), indicating the great advantage of binarizing the real-valued parameters. Moreover, CCCF shows superior retrieval time compared to DCF while enjoying a better accuracy. Thus, CCCF is a suitable model for large-scale recommender systems where the retrieval time is restricted within a limited time quota.

\subsubsection{\bf Impact of Hyper-parameter $G$ and $r$ (RQ2)}
\label{rq2}

Our CCCF has two key parameters, the number of components $G$ and code length $r$, to control the complexity and capacity of CCCF. Figure \ref{hyper1} and \ref{hyper2} evaluate how they affect the recommendation performance under varied total bits and fixed total bits. In Figure \ref{hyper1}, we vary the code length $r$ from $4$ to $16$ and the component number $G$ from $1$ to $16$. We can see that increasing $G$ leads to continued improvements. When $G$ is larger than $8$, the improvement tends to become saturated as the number of components increases. In addition, a larger value of $G$ would lead to relatively longer training time. Similar observations can be found from the results of hyper-parameter $r$ evaluation. In Figure \ref{hyper2}, we fix the total bits $rG$ in range $\{32,64,96,128\}$ and varies the component number $G$ from $1$ to $32$. It should be noted that when $G=1$, CCCF model is identical to DCF model. As we gradually increase component number $G$, the recommendation performance grows since the real-valued component weight could enhance the representation capability. The best recommendation performance is achieved when $G=8$ or $16$. When $G$ is larger than the optimal values, increasing $G$ will hurt the performance. The main reason is that we fix the total bits $rG$, so that larger values of $G$ lead to smaller values of $r$. This will reduce the learning space of CCCF since the component weight is calculated by a predefined distance function which does not consider the rating information.

\vspace{-2pt}
\subsubsection{\bf Impact of sparsity of component weight and integer weight scaling (RQ3)}
\label{rq3}

\begin{figure}[h]
  \centering
    \includegraphics[width=0.23\textwidth]{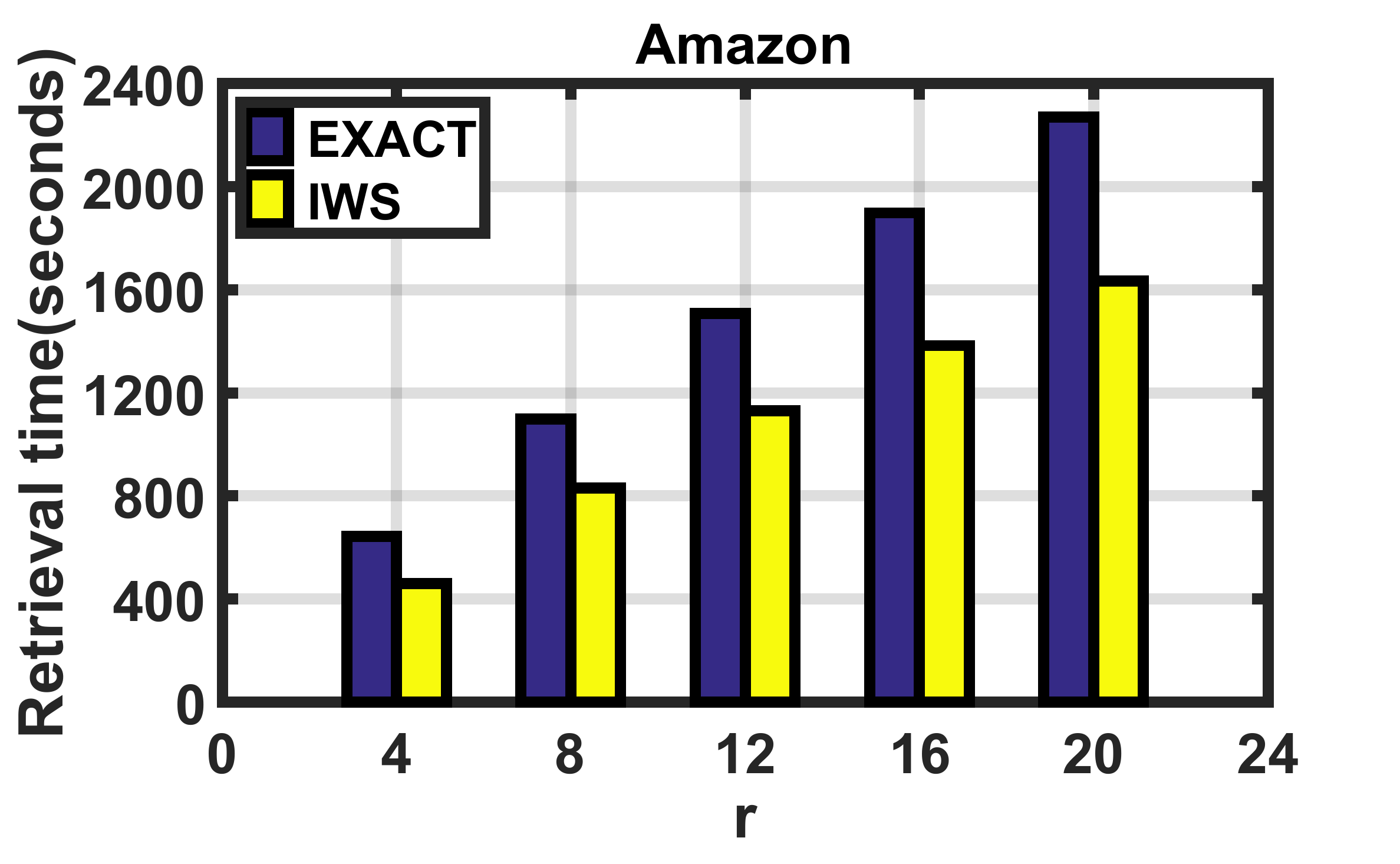}
    \includegraphics[width=0.23\textwidth]{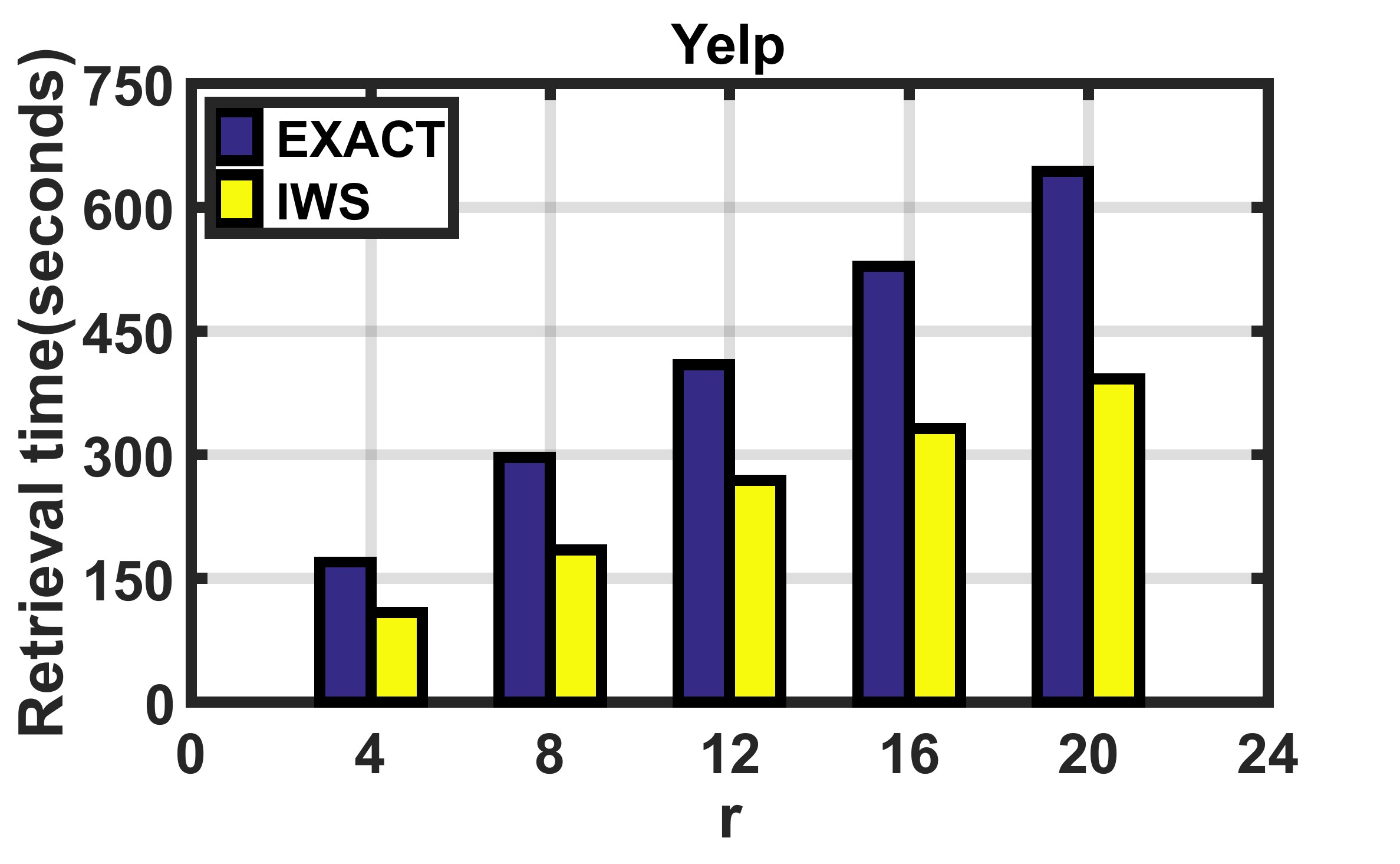}
  \caption{Retrieval cost of the naive version and IWS version.}
\label{retrieval_e}
\end{figure}

To demonstrate the effectiveness of the integer weight scaling in accuracy and retrieval time , we run two versions of CCCF: EXACT and IWS.  EXACT does not adopt the proposed integer weight scaling strategy. IWS uses the integer weight scaling described in Section \ref{si}. Figure \ref{retrieval_e} summarizes the speedup of the two versions of CCCF. The retrieval cost is not  sensitive to the integer scaling parameter $e$ and we set $e$ to $100$. We can see the gap between the versions is consistent over Amazon and Yelp dataset. The low cost of IWS validates the effectiveness of the integer scaling which replaces the floating-point operations with the fast integer computation.

\begin{figure}[h]
  \centering
    \includegraphics[width=0.23\textwidth]{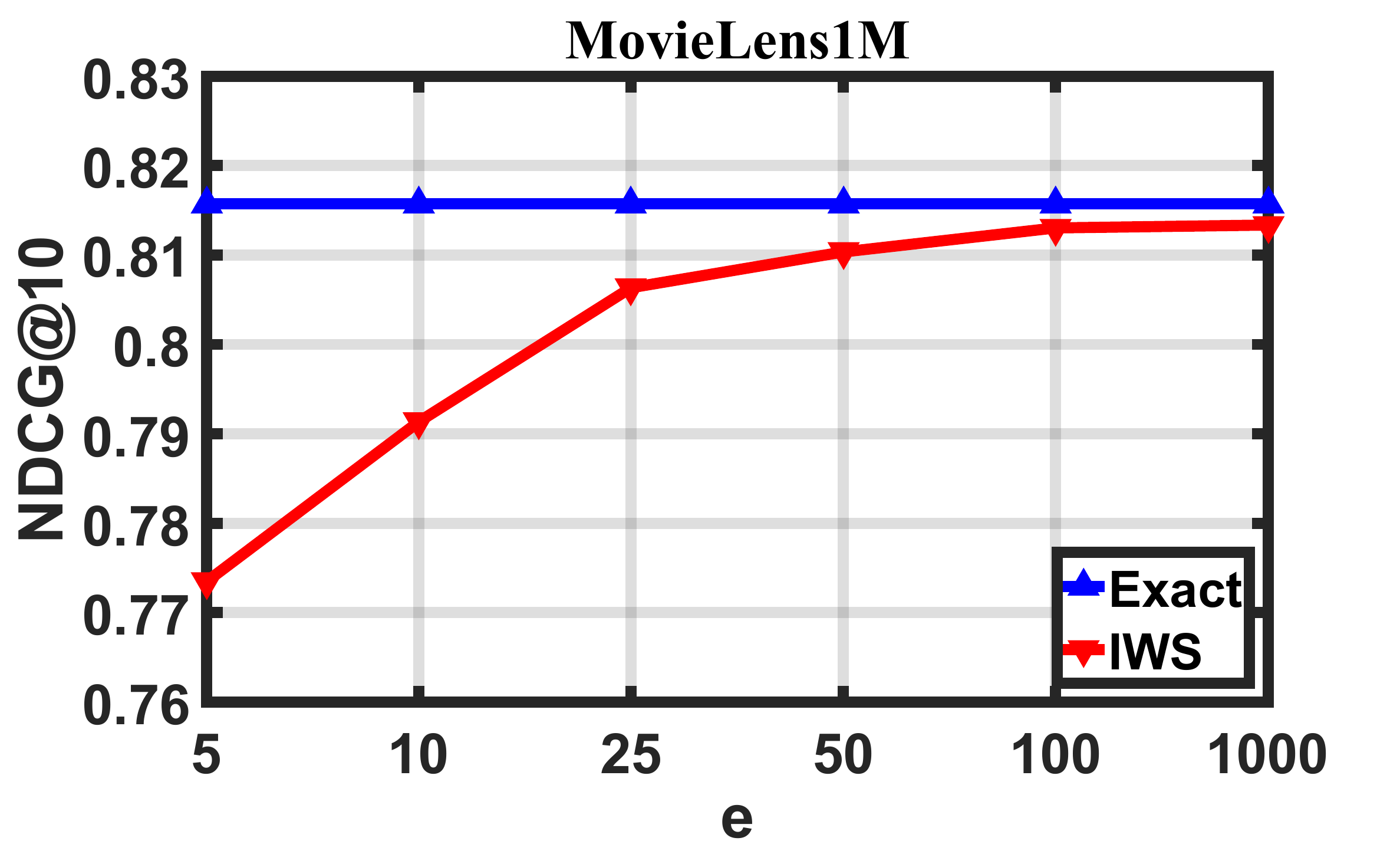}
    \includegraphics[width=0.23\textwidth]{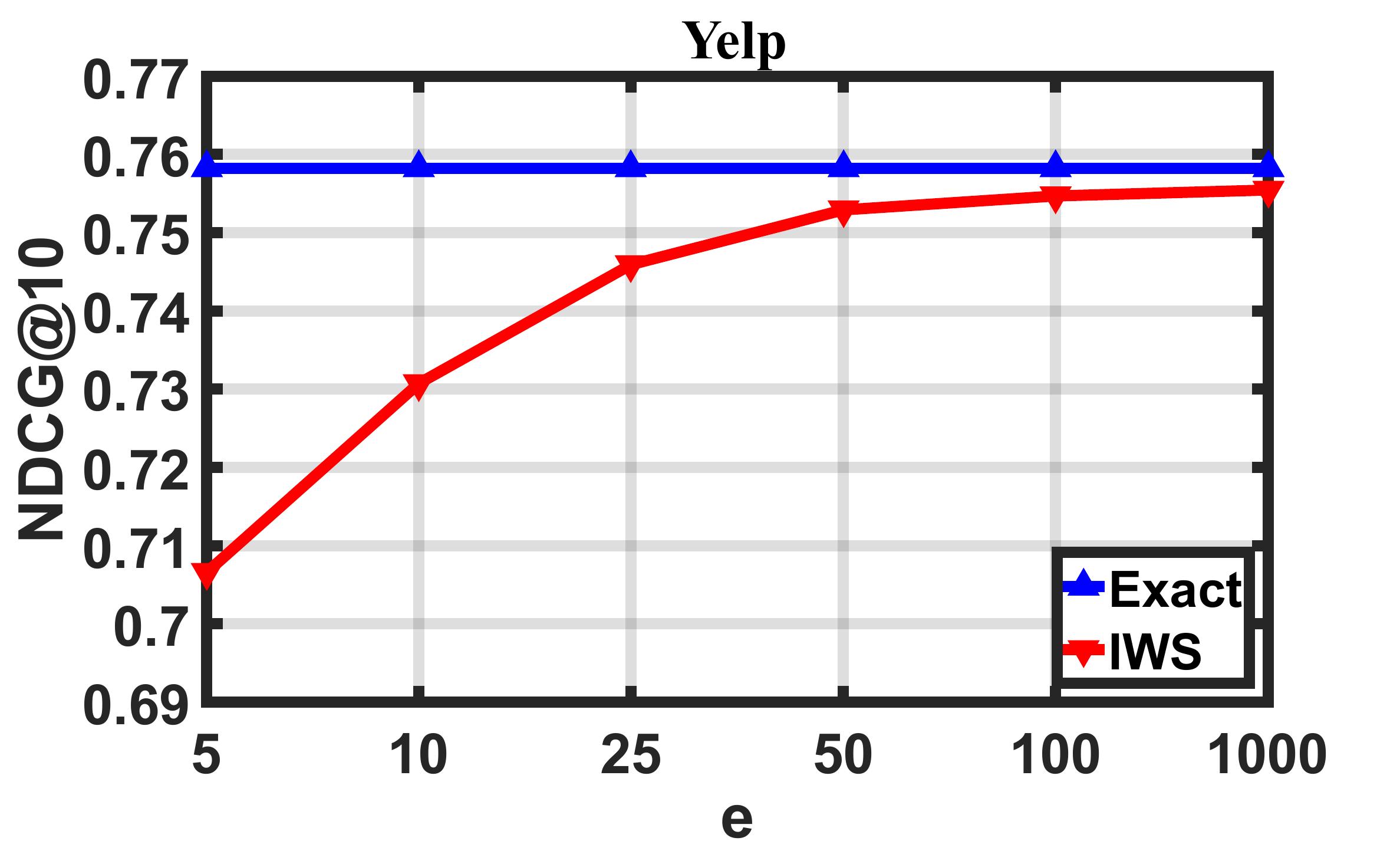}
    \vspace{-0.1cm}
  \caption{Performance of CCCF with different $e$ values.}
\label{acc_e}
\end{figure}

Figure \ref{acc_e} shows the impact of hyper-parameter $e$ on the accuracy of CCCF. We can find that when $e$ is smaller than $100$, the increase of  $e$ leads to gradual improvements. When $e$ is larger than $100$, further increasing its value cannot improve the performance. This indicates that IWS is relatively insensitive when $e$ is sufficiently large. We thus suggest to set $e$ to 100.

\begin{figure}[h]
  \centering
    \includegraphics[width=0.23\textwidth]{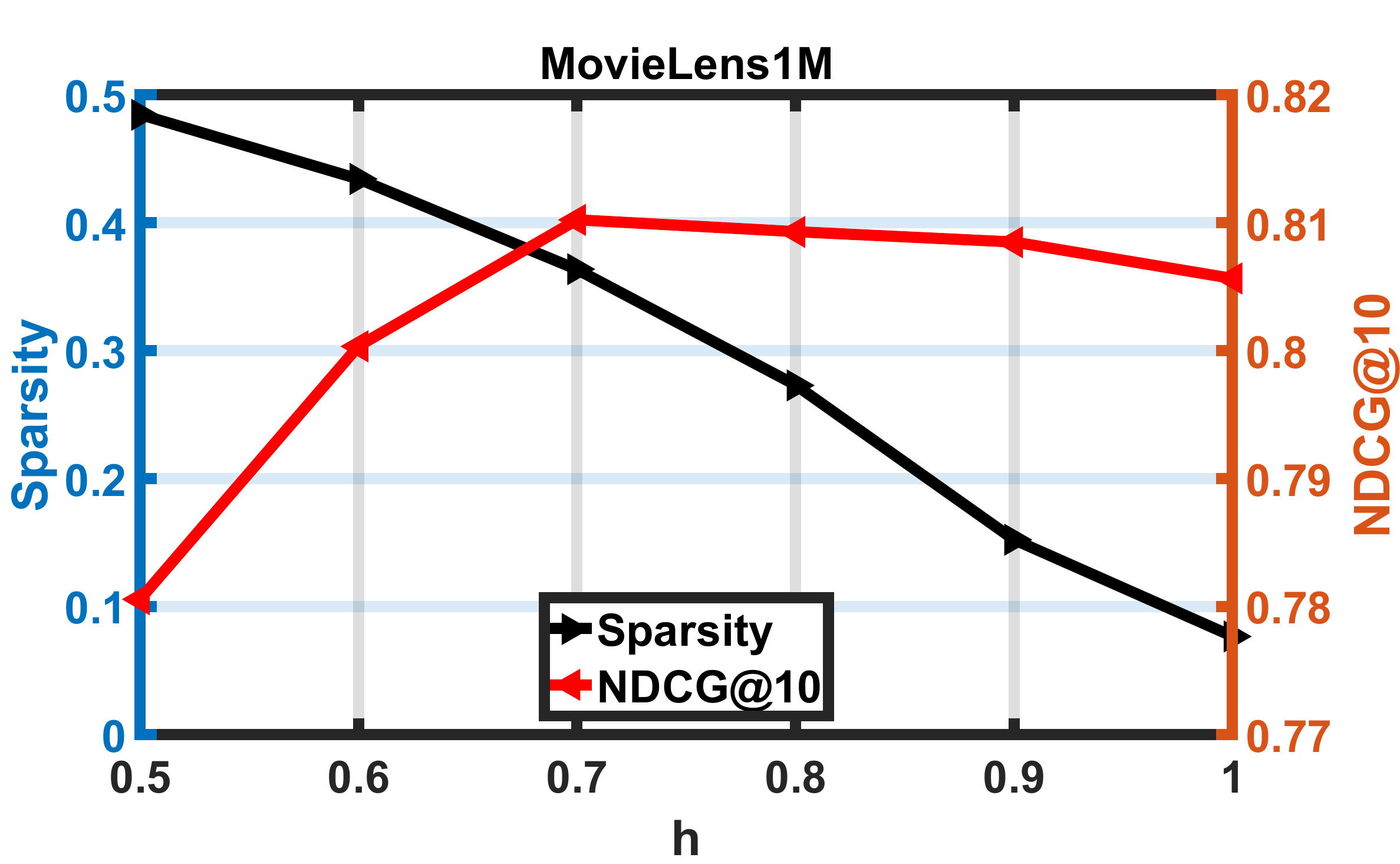}
    \includegraphics[width=0.23\textwidth]{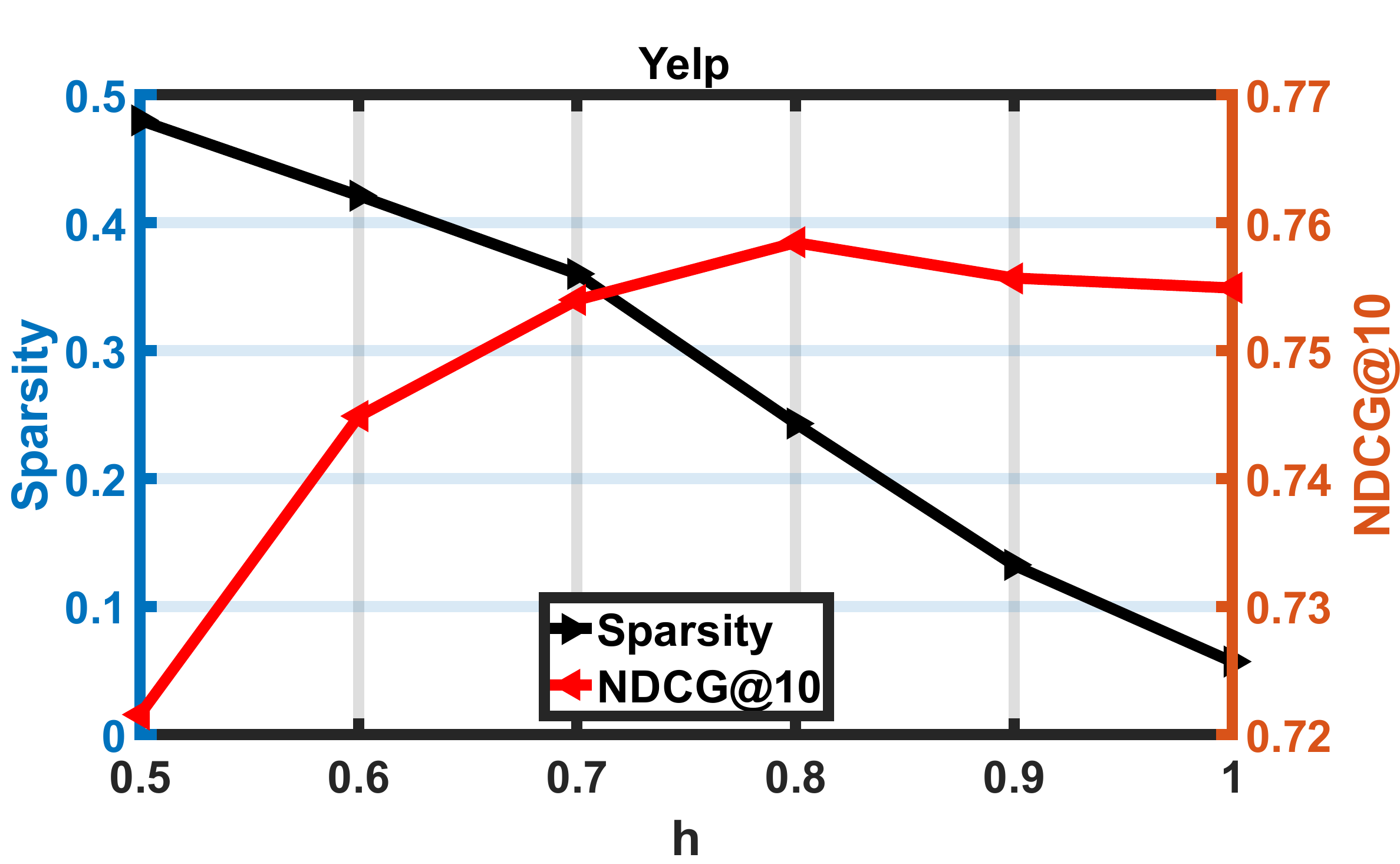}
        \includegraphics[width=0.23\textwidth]{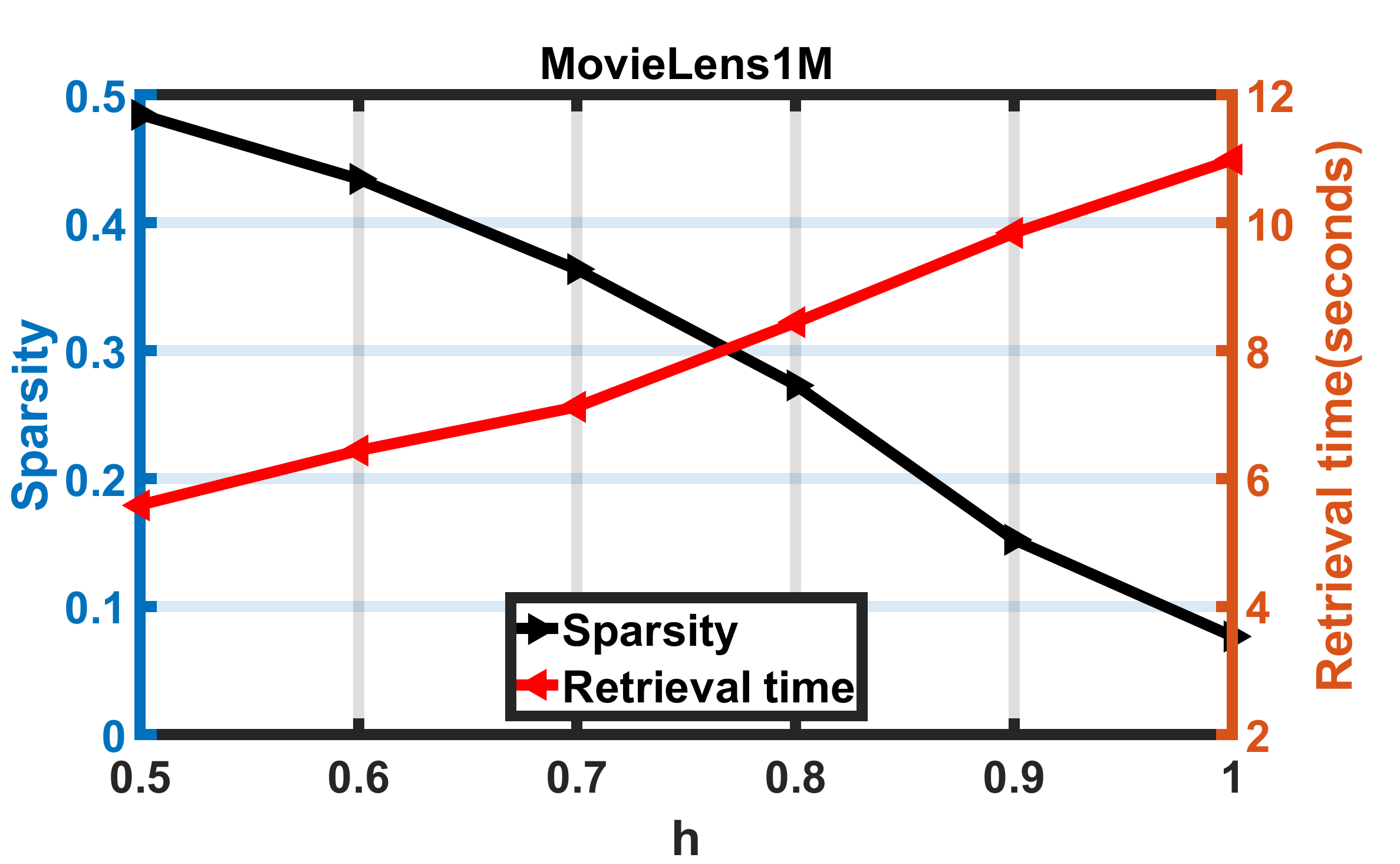}
    \includegraphics[width=0.23\textwidth]{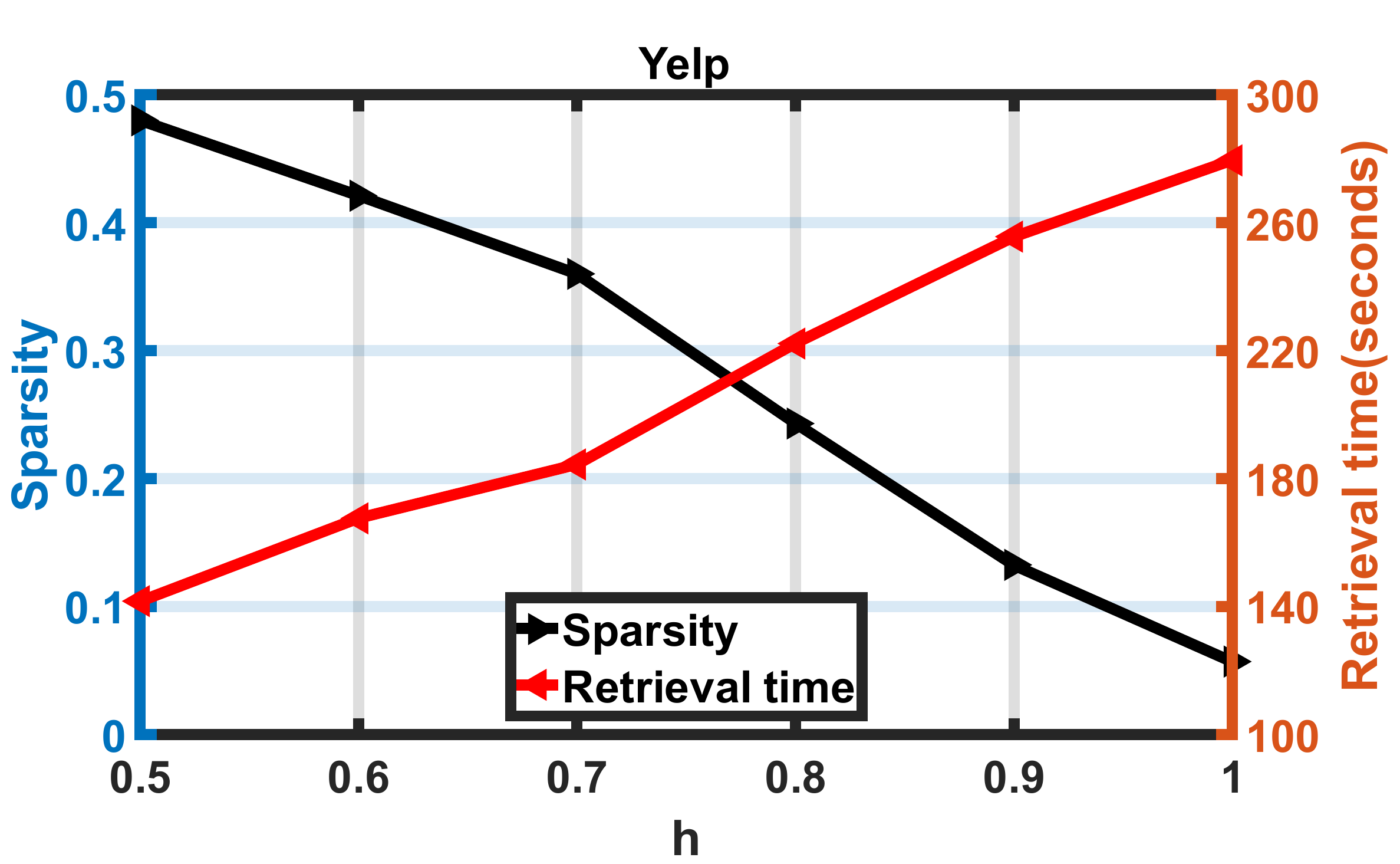}
  \caption{Item recommendation performance and retrieval cost of CCCF with different $h$ values (controlling the sparsity of user/item component weights).}
\label{hyper_h}
\end{figure}

To reveal the impact of sparsity of component weight in accuracy and retrieval cost, we vary the bandwidth hyperparameter $h$ from $0.5$ to $1$. It is obvious that decreasing the value of $h$ will increasing the sparsity of component weights. Figure \ref{hyper_h} shows the accuracy and retrieval cost of CCCF for differnt $h$. First, we can see that the retrieval cost of CCCF continuously drop as we decrease the values of $h$ since high sparsity lead to fast computation. Second, we observe that the best recomendation performance is achieved when $h=0.7\sim0.8$. When $h$ is smaller than $0.7\sim0.8$, increasing the sparsity will make CCCF robust to overfitting. However, when the sparsity level is quite high, the CCF model might not be informative enough for prediction and thus suffer performance drop.

\subsubsection{\bf Item Embeddings Visualization (RQ4)}
\begin{figure}[h]
  \centering
  \hspace{-0.2cm}
    \includegraphics[width=0.24\textwidth]{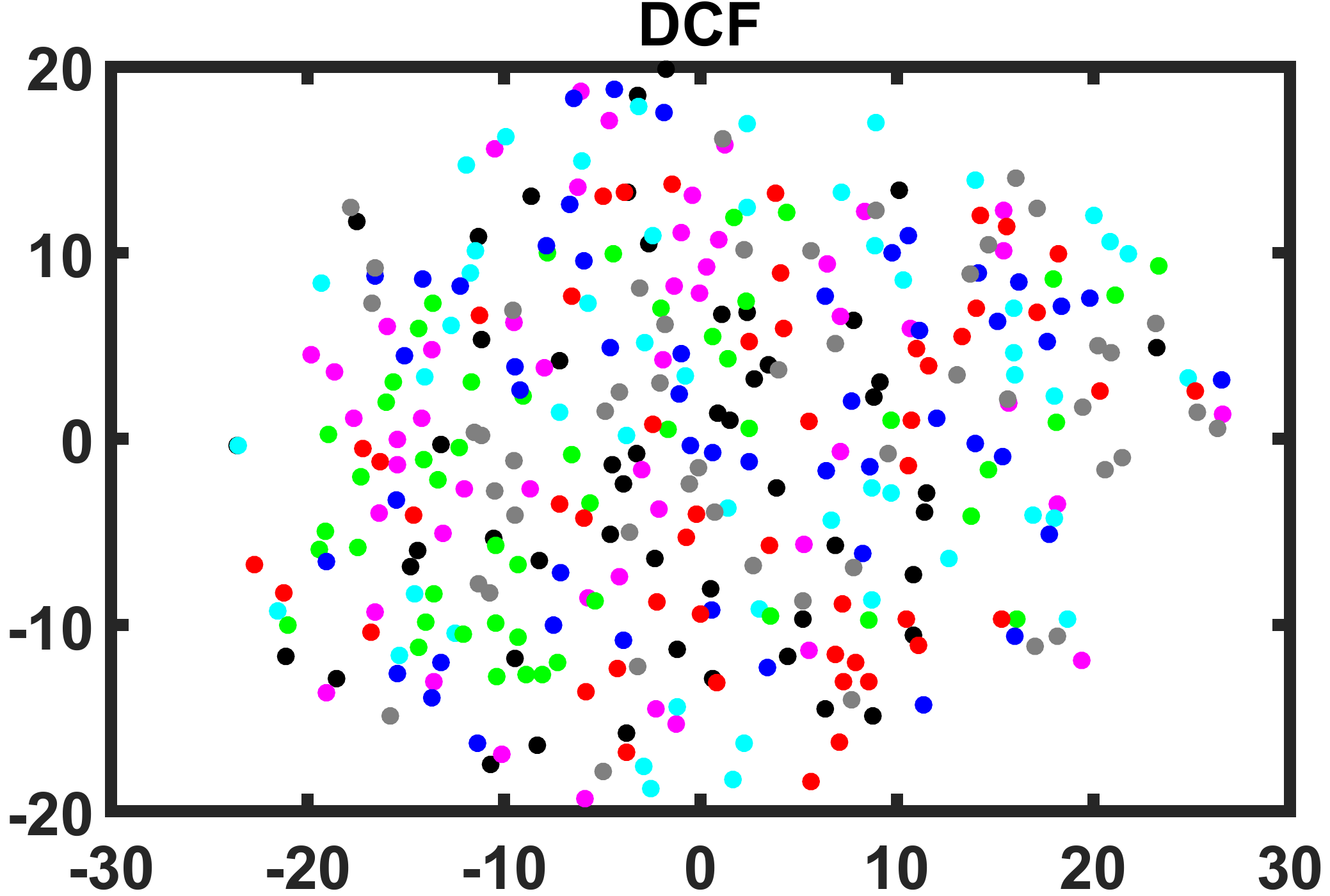}
    \hspace{-0.18cm}
    \includegraphics[width=0.24\textwidth,height=0.163\textwidth]{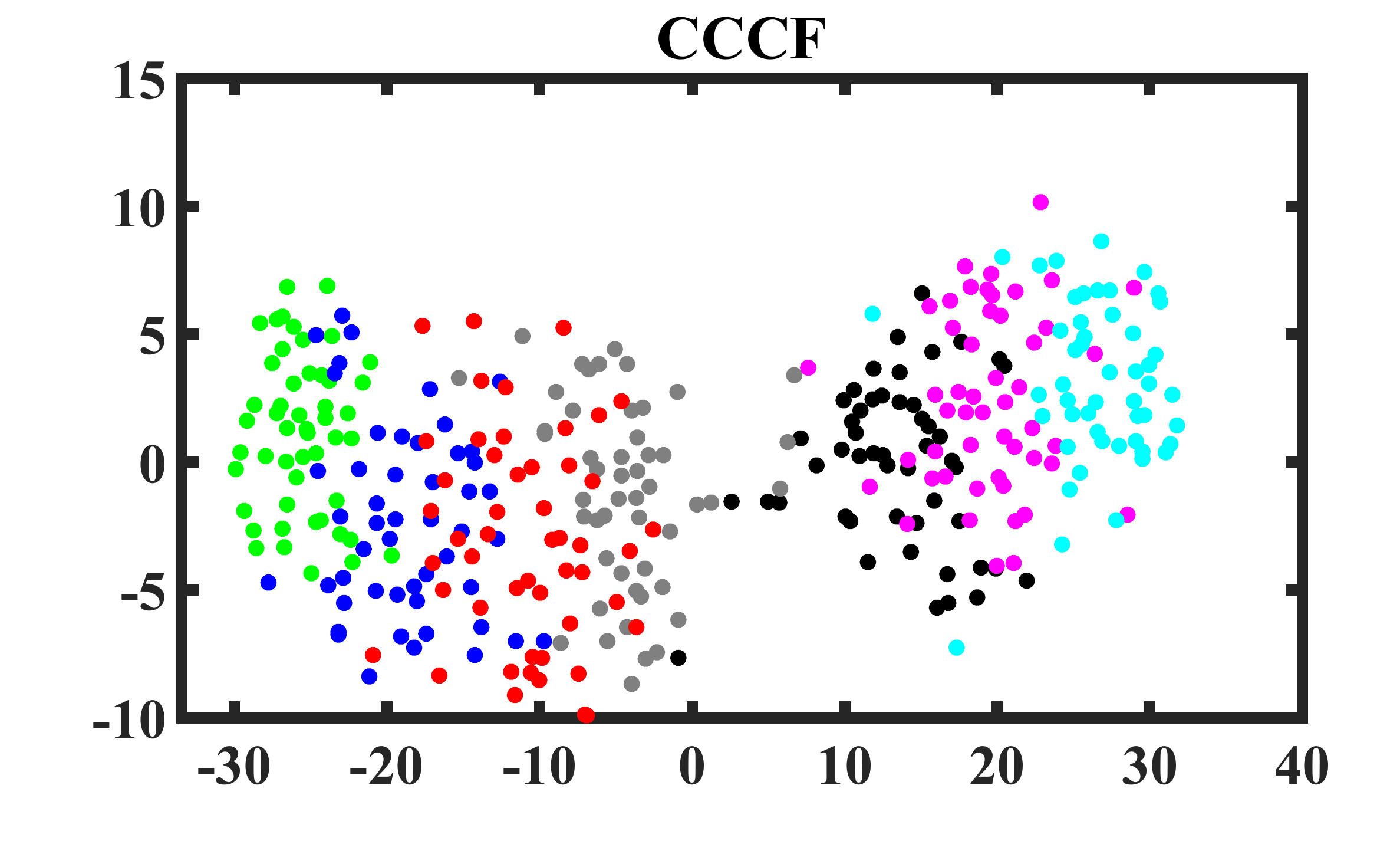}
  \caption{Visualization of Moivelens 1M dataset. Each point indicates one item embedding (movie). The color of a point indicates the genre of the movie.}
\label{vis}
\end{figure}

The key advantage of compositional codes is the stronger representational capability in comparison to binary codes. Therefore we visualize the learned item embeddings of the movielens 1M dataset where items are indicated as movies. We use the item representations learned by DCF and CCCF as the input to the visualization tool t-SNE \cite{maaten2008visualizing}. As a result, each movie is  mapped into a two-dimensional vector. Then we can visualize each item embedding as a point on a two dimensional space. For items which are labelled as different genres, we adopt different colors on the corresponding points. Thus, a good visualization result is that the points of the same color are closer to each other. The visualization result is shown in Figure \ref{vis}. We can find that the result of DCF is unsatisfactory since the points belonging to different categories are mixed with each other. For CCCF, we can observed clear clusters of different categories. This again validates the advantage of much stronger representation power of the compositional codes over traditional binary codes.

\section{Related Work}
As a pioneer work, Locality-Sensitive Hashing has been adopted for generating hash codes for Google News readers based on their click history \cite{das2007google}. Following this work, random projection was applied for mapping  learned user/item embeddings from matrix factorization into the Hamming space to obtain binary codes for users and items \cite{karatzoglou2010collaborative}. Similar to the idea of projection, Zhou et al. \cite{zhou2012learning} generated binary codes from rotated continuous user/item representations by running Iterative Quantization. In order to derive more compact binary codes, the de-correlated constraint over different binary codes was imposed on user/item continuous representations and then rounded them to produce binary codes \cite{liu2014collaborative}. The relevant work could be summarized as two independent stages: relaxed learning of user/item representations with some specific constraints and subsequent bianry quantization. However, such two-stage approaches suffer from a large quantization loss according to \cite{zhang2016discrete}, so direct optimization of matrix factorization with discrete constraints was proposed. To derive compact yet informative binary codes, the balanced and de-correlated constraints were further imposed \cite{zhang2016discrete}. In order to incorporate content information from users and items, content-aware matrix factorization and factorization machine with binary constraints was further proposed \cite{lian2017discrete,liu2018discrete}. For dealing with social information, a discrete social recommendation model was proposed in \cite{liu2019discrete}.

Recently, the idea of compositional codes has been explored in the compression of feature embedding \cite{chen2018learning,shu2017compressing,svenstrup2017hash}, which has become more and more important in order to deploy large models to small mobile devices. In general, they composed the embedding vectors using a small set of basis vectors. The selection of basis vectors was governed by the hash code of the original symbols. In this way, compositional coding approaches could maximize the storage efficiency by eliminating the redundancy inherent in representing similar
symbols with independent embeddings. In contrast, this work employs compositional codes to address the inner product similarity search problem in recommender systems.

\section{Conclusion and Future Work}
This work contributes a novel and much more effective framework called Compositional Coding for Collaborative Filtering (CCCF). The idea is to represent each user/item by multiple components of binary codes together with a sparse weight vector, where each element of the weight vector encodes the importance of the corresponding component of binary codes to the
user/item. In contrast to standard binary codes, compositional codes significantly enriches the representation capability without sacrificing retrieval efficiency. To this end, CCCF can enjoy both the merits of effectiveness and efficiency in recommendation. Extensive experiments demonstrate that CCCF not only outperforms existing hashing-based binary code learning algorithms in terms of recommendation accuracy, but also achieves considerable speedup of retrieval efficiency over the state-of-the-art binary coding approaches. In future, we will apply compositional coding framework to other recommendation models, especially for the more generic feature-based models like Factorization Machines. In addition, we are interested in employing CCCF on the recently developed neural CF models to further advance the performance of item recommendation.
\begin{acks}
This research is supported by the National Research Foundation Singapore under its AI Singapore Programme [AISG-RP-2018-001]. Xin Wang is supported by China Postdoctoral Science Foundation No. BX201700136. 
\end{acks}
\newpage
\bibliographystyle{ACM-Reference-Format}
\bibliography{references}

\end{document}